\documentclass[sigconf,natbib=true]{acmart}
\AtBeginDocument{%
  \providecommand\BibTeX{{%
    \normalfont B\kern-0.5em{\scshape i\kern-0.25em b}\kern-0.8em\TeX}}}

\setcopyright{acmcopyright}
\copyrightyear{2018}
\acmYear{2018}
\acmDOI{XXXXXXX.XXXXXXX}

\acmConference[Conference acronym 'XX]{Make sure to enter the correct
  conference title from your rights confirmation emai}{June 03--05,
  2018}{Woodstock, NY}
\usepackage{xspace}
\usepackage{amsthm}
\usepackage{amsmath}
\usepackage{enumitem}
\usepackage{booktabs}
\usepackage{multirow}
\usepackage{graphicx}
\usepackage[normalem]{ulem}
\useunder{\uline}{\ul}{}
\usepackage{float}
\usepackage{caption}
\usepackage{subcaption}
\usepackage{balance}
\usepackage[toc,page]{appendix}

\newcommand{\modelname}{\textsf{MPKG}\xspace}
\acmPrice{15.00}
\acmISBN{978-1-4503-XXXX-X/18/06}

\begin{document}

\title{Zero-shot Item-based Recommendation via Multi-task Product Knowledge Graph Pre-Training}

\author{Ziwei Fan}
\authornote{Work done during the internship at Salesforce AI Research.}
\email{zfan20@uic.edu}
\affiliation{%
  \institution{University of Illinois Chicago}
  \city{Chicago}
  \state{IL}
  \country{USA}
}

\author{Zhiwei~Liu}
\email{zhiweiliu@salesforce.com}
\affiliation{%
  \institution{Salesforce AI Research}
  \streetaddress{181 Lytton Ave}
  \city{Palo Alto}
  \state{CA}
  \country{USA}
  \postcode{94301}
}

\author{Shelby~Heinecke}
\email{shelby.heinecke@salesforce.com}
\affiliation{%
  \institution{Salesforce AI Research}
  \streetaddress{181 Lytton Ave}
  \city{Palo Alto}
  \state{CA}
  \country{USA}
  \postcode{94301}
}

\author{Jianguo Zhang}
\email{jianguozhang@salesforce.com}
\affiliation{%
  \institution{Salesforce AI Research}
  \streetaddress{181 Lytton Ave}
  \city{Palo Alto}
  \state{CA}
  \country{USA}
  \postcode{94301}
}

\author{Huan~Wang}
\email{huan.wang@salesforce.com}
\affiliation{%
  \institution{Salesforce AI Research}
  \streetaddress{181 Lytton Ave}
  \city{Palo Alto}
  \state{CA}
  \country{USA}
  \postcode{94301}
}

\author{Caiming~Xiong}
\email{cxiong@salesforce.com}
\affiliation{%
  \institution{Salesforce AI Research}
  \streetaddress{181 Lytton Ave}
  \city{Palo Alto}
  \state{CA}
  \country{USA}
  \postcode{94301}
}

\author{Philip S. Yu}
\email{psyu@uic.edu}
\affiliation{%
  \institution{University of Illinois Chicago}
  \city{Chicago}
  \state{IL}
  \country{USA}
}

\renewcommand{\shortauthors}{Trovato and Tobin, et al.}

\begin{abstract} 
Existing recommender systems face difficulties with zero-shot items, \textit{i.e.} items that have no historical interactions with users during the training stage. 
Though recent works extract universal item representation via pre-trained language models (PLMs), they ignore the crucial item relationships.  
This paper presents a novel paradigm for the Zero-Shot Item-based Recommendation (ZSIR) task,
which pre-trains a model on product knowledge graph (PKG) to refine the item features from PLMs.
We identify three challenges for pre-training PKG, which are multi-type relations in PKG, semantic divergence between item generic information and relations and domain discrepancy from PKG to downstream ZSIR task. 
We address the challenges by proposing four pre-training tasks and novel task-oriented adaptation (ToA) layers.
Moreover, this paper discusses how to fine-tune the model on new recommendation task such that the ToA layers are adapted to ZSIR task.
Comprehensive experiments on $18$ markets dataset are conducted to verify the effectiveness of the proposed model in both knowledge prediction and ZSIR task.
\end{abstract}


\maketitle

\section{Introduction}
\begin{figure}[]
\centering
\includegraphics[width=0.45\textwidth]{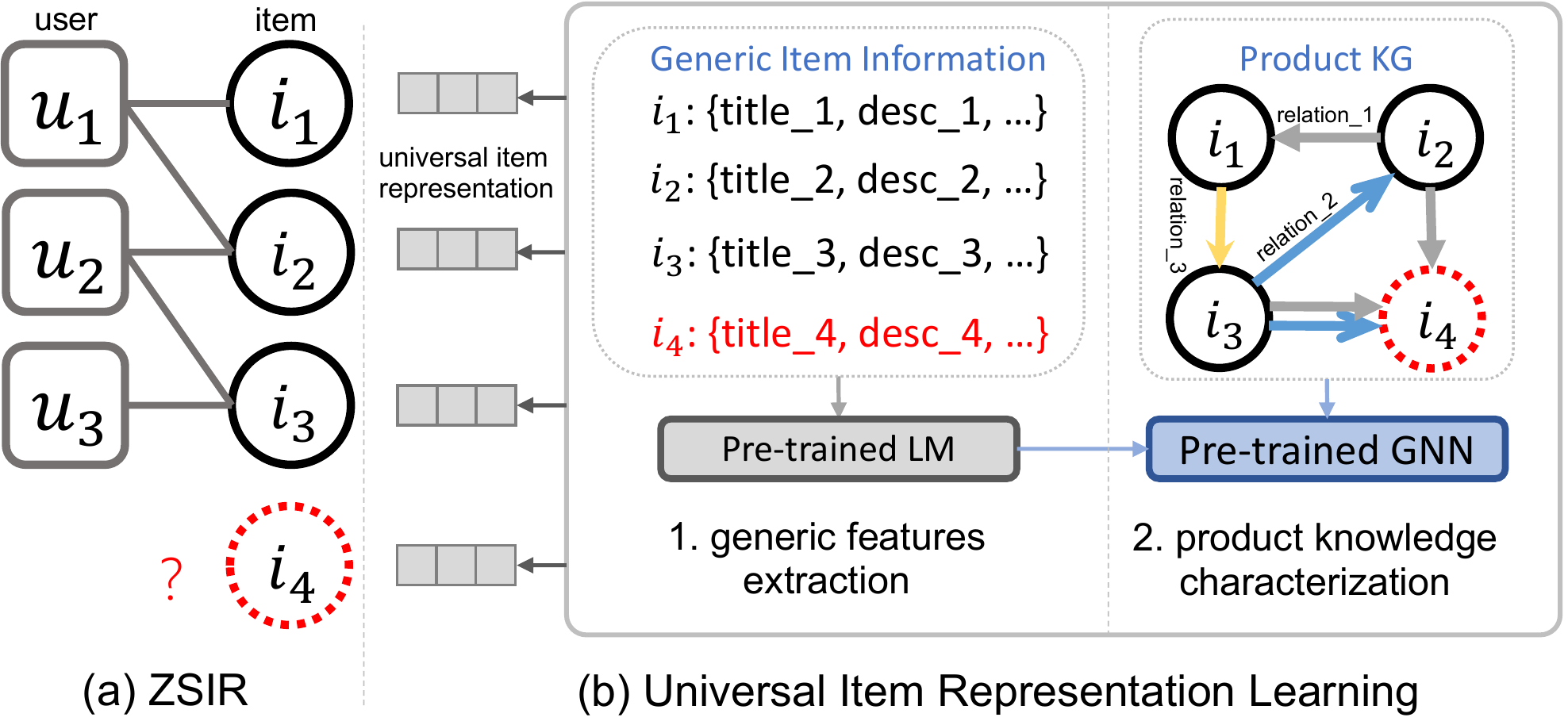}
\caption{(a) A toy example of Zero-shot Item-based Recommendation (ZSIR), where the zero-shot item $i_4$ has no interaction with users. (b) A general framework for universal item representation learning, which contains universal feature extraction and product knowledge characterization. }
\label{fig:toy_example}
\end{figure}

Recommender systems~(RS) provide personalized information retrieval services to users, and have increasingly become an irreplaceable component in several web applications, such as fashion~\cite{chen2019personalized} and movies~\cite{diao2014jointly}. 
Most existing collaborative filtering methods~\cite{wang2019neural,ferrari2019we,he2020lightgcn} leverage collaborative signals from the historical interactions of users and items.
However, collaborative filtering methods are unable to resolve the item cold-start problem. In the item cold-start problem, some items have few to no historical interactions~\cite{li2019zero,huan2022industrial}. Without historical iteractions, the representations of cold-start items are not optimized during collaborative filtering training. 
In this paper, we tackle the Zero-Shot Item-based Recommendation task~(ZSIR). 
We present a toy example of ZSIR in Figure~\ref{fig:toy_example}(a). Compared with other items, $i_4$ has no interactions with users, and hence $i_4$ is a zero-shot item. 

Item-based recommendation methods~\cite{xue2019deep, ning2012sparse, he2018nais} represent users as weighted sum of interacted items.
The key to resolve the ZSIR task becomes learning representations for zero-shot items. 
Because of the unavailability of interaction data, we believe the foundation is to infer embeddings of zero-shot items from generic side information, such as \textit{titles} and \textit{descriptions}. 
With the booming of large pre-trained language models~(PLMs)~\cite{devlin2018bert,sanh2019distilbert,reimers2019sentence,raffel2020exploring}, powerful tools are available to infer universal item representations from textual data. 
PLMs have been explored in recent recommender systems~\cite{hou2022towards,geng2022recommendation}. In these works, PLMs are used to infer item representations, and these item representations are then used as input for downstream recommendation tasks. 
Nevertheless, we argue that direct inference of item representations from PLMs is far from aligning the semantics of items for recommendation, thus impairing the ZSIR performance. 
High-quality universal item representations should not only capture the semantics of generic information, but should also incorporate recommendation-oriented knowledge~\cite{xu2020knowledge,xu2020inductive}, such as complementary and substitution relationships among items.

To this end, we propose a novel universal item representation learning framework, which comprises two components, \textit{i.e.} generic features extraction and product\footnote{The terms ``product'' and ``item'' are used interchangeably.} knowledge characterization.
The illustration of our universal item representation learning process is presented in Figure~\ref{fig:toy_example}(b).
The generic features extraction module employs a PLM to extract features from generic item side information, such as \textit{titles}, \textit{descriptions}, etc. 

However, features directly extracted from generic information are hard to adapt to recommendation task. 
Therefore, we propose the product knowledge characterization module to enhance the universal representation of items for recommendation.
To be more specific, we construct the product knowledge graph~(PKG) to represent recommendation-oriented knowledge, where nodes are items and edges are different relations between items, such as complementary, substitution, and etc. 
Since those relations in PKG are usually retrieved from user-item interactions~\cite{xu2020product}, leveraging PKG for refinement adapts the universal representation for recommendation.
We pre-train a graph neural network (GNN) model~\cite{hu2020gpt,qiu2020gcc} to refine the features extracted from PLM such that the final universal item representations capture semantics relevant to recommendation tasks.
It is noteworthy that we ensure the pre-trained GNN has the inductive ability for zero-shot items, since those items may not present in the pre-training stage of the GNN model.
We demonstrate a toy example of the PKG over items in Figure~\ref{fig:toy_example}(b), which 
contains three item-item relationships.



The challenges of pre-training PKG are from the following three perspectives: 
1) Multi-type Relations in PKG; 
2) Semantic Divergence between generic information and relations; 
3) Domain Discrepancy from PKG to downstream ZSIR task. 

Firstly, the multi-type relations intrinsically exist in PKG due to various item-item relationships.  
Pre-training a PKG encoder demands comprehensive characterization of those multiple relations, which is still under-explored.
Secondly, the universal features are extracted from generic information, while the relations in PKG may reflect different semantics. 
For example, two items has similar \textit{titles}, but they have no complementary relations.
Ignoring the semantic divergence disables the unifying ability of graph encoder to incorporate both generic information and relation semantics. 
Thirdly, we use a pre-trained encoder to inference item embeddings for ZSIR. 
However, because the domain of ZSIR task is distinct from PKG domain, though PKG yield similar embeddings for two items, users may have different preferences towards them in ZSIR task. 
Moreover, multiple relations may have different contributions to ZSIR task from encoding the PKG. 
We refer this as the domain discrepancy issue.
To resolve aforementioned problems, we
propose a novel \textbf{M}ulti-task \textbf{P}roduct \textbf{K}nowledge \textbf{G}raph model for pre-training, and devise a novel paradigm to fine-tune the pre-trained model on recommendation task. 
\modelname is able to adapt to the downstream recommendation task and infer the universal representation of zero-shot items.
To be more specific, we extract multiple single-relation PKG from the original PKG and adopt the SGCN~\cite{wu2019simplifying} as the encoder for each single-relation PKG. 
SGCN is advantageous due to its light weights and inductive inference ability. 
Then, we endow this multi-relation graph encoder with adaptation ability to different tasks via novel Task-oriented Adaptation (ToA) layers.
ToA layers intake embeddings from graph encoder, and output task-oriented embeddings.
We also devise four pre-training tasks to optimize the graph encoder and ToA layers, which are Knowledge Reconstruction (KR), High-order Neighbor Reconstruction (HNR), universal Feature Reconstruction (FR), and Meta Relation Adaptation (MRA) tasks. 
KR and HNR tasks function together to characterize the multi-type relations in PKG. 
HNR task aims at alleviating semantic divergence problem.
And, MRA task targets at adapting the pre-trained models to ZSIR task during the fine-tuning stage.
Each task is associated with one type of ToA layer. 
Experiments are conducted on the cross-market dataset~\cite{bonab2021crossmarket}, which consists of $18$ different markets data from Amazon. 
We summarize our contributions as follows:
\begin{itemize}[leftmargin=*]
    \item We propose a novel PKG pre-training and fine-tuning framework to tackle the ZSIR problem, which enhances the \modelname with inductive ability.
    \item We identify three type of challenges in PKG pre-training and devise four pre-training tasks. The MRA pre-training task is firstly proposed for adapting model to new downstream tasks.
    \item We propose a novel task-oriented adaption layer for each task, which adapts the embeddings from multi-relation graph encoder to different tasks. 
    \item We conduct comprehensive experiments on 18 different markets to verify the effectiveness of our \modelname and its extrodinary ability in solving ZSIR. 
\end{itemize}

\section{Related Work}

\subsection{Product Knowledge Graph}
Product knowledge graphs~(PKGs) are graphs whose nodes represent products (items) and whose edges represent various item-item relationships, such as complementary and substitute relationships. The item-item relationships are extracted from meta data of items or interaction data on items. Existing literature of PKG for recommendation mainly focuses on two directions, including PKG construction~\cite{dong2020autoknow, xu2020product, mcauley2015inferring} and PKG utilization for recommendation~\cite{zhao2017improving, yan2022personalized, zhou2022decoupled, xu2020knowledge, chen2020try, wang2020make}. 

Regarding the PKG construction, the earliest work is Sceptre~\cite{mcauley2015inferring}, which focuses on modeling product complementary and substitution relationships by using topic modeling methods on product review data. \cite{xu2020product} further introduces three more knowledge relationships between products, including co-view and IsA. Moreover, Autoknow~\cite{dong2020autoknow} builds the PKG from item textual knowledge and user-item interaction data to encode thousands of relation types. Those works suggest that the item relationships described in PKG are more closely related to item semantics in recommendation~\cite{xu2020product,dong2020autoknow}. 

Utilizing PKG for recommendation also attracts increasing attention. RSC~\cite{zhao2017improving} proposed complement and substitution networks to improve rating prediction accuracy, demonstrating the effectiveness of additional item relationships. \cite{xu2020knowledge} developed Bayesian dual embedding framework to encode complementary item relationships for recommendation. Chorus~\cite{wang2020make} encoded item relationships into sequential recommendation with temporal kernel functions.
In summary, constructing and leveraging PKGs is a promising direction for improving recommendation performance. 
In this work, we firstly propose to pre-train a PKG model and then fine-tune this model in recommendation task.


\subsection{Graph Pre-Training}
Graph neural networks~(GNNs) encode rich relationships between nodes and formulates these connections as a graph. As a great amount of data can be represented as a graph, the GNN representation learning and its pre-training become an important but challenging research problem. 
Several classical GNN pre-training methods were developed for general graph tasks.
GPT-GNN~\cite{hu2020gpt} proposed two pre-training generation tasks, including node attributes generation and edge generation. The demonstration of GPT-GNN is conducted in downstream tasks with time and data shifts. Another representative work GCC~\cite{qiu2020gcc} assumes that the graph structural property is transferable and universal across different networks. The authors defined the $r$-ego network as the positive subgraph and proposed the negative subgraph sampling into the contrastive learning.  \cite{lu2021learning} proposed to bridge the gap between graph pre-training and fine-tuning with model-agnostic meta-learning strategy. It focused on subgraph learning node-level embedding learning for predicting the node connections, subgraph graph-level learning for predicting how close between the subgraph and the whole graph. Each subgraph is used as the support set for node and graph levels adapation meta-learning. 

These GNN pre-training methods all demonstrate the effectiveness of GNN in generalization to unseen tasks and nodes learning. Moreover, several works about inductive graph learning~\cite{hamilton2017inductive,xu2020inductive} also show the superiority of GNN to inductively adapt to infer unseen nodes. In this work, we pre-train a GNN using a PKG.

\section{Preliminaries}
Our work is focused on strategically pre-training a graph neural network for PKG. We begin with the definition of PKG.
\begin{definition}
\textbf{Product Knowledge Graph (PKG).} A product knowledge graph~(PKG) is denoted as $\mathcal{G}=\{\mathcal{I}, \mathcal{E}, \mathcal{R}, \mathbf{X}, \theta\}$, where $\mathcal{I}$ and $\mathcal{E}$ denote the sets of item nodes and edges, respectively. 
$\mathcal{R}$ is the relation type of edges, which is associated with $\mathcal{E}$ via a edge-type mapping function $\theta: \mathcal{E}\to \mathcal{R}$. $\mathbf{X}\in \mathbb{R}^{|\mathcal{I}|\times d}$ denotes the feature vector for nodes, which is extracted from item generic side information via PLMs. For each edge type $r\in\mathcal{R}$, we define its \textbf{$r$-PKG} as $\mathcal{G}^r=\{\mathcal{I}, \mathcal{E}^r, \mathbf{X}\}$, where $\mathcal{E}^r$ only has edges in relation $r$. 
\end{definition}
Note that PKG only has one node type, \textit{i.e.} items, and usually has multiple edge types (relations) between items, \textit{e.g.} \textit{co-purchasing}, \textit{co-view}, etc. 
To achieve knowledge-enhanced universal item representations, we pre-train a model that encodes nodes to embeddings in PKG.  
\begin{definition}
\textbf{PKG Pre-training.} Given a PKG $\mathcal{G}$, the PKG pre-training task is to learn an encoder \text{Enc}($\mathcal{G}$)$\rightarrow \mathbf{E} \in \mathbb{R}^{|\mathcal{I}|\times d}$, where each node $i\in \mathcal{I}$ is represented as an embedding $\mathbf{e}_{i} \in \mathbb{R}^{d}$.
\end{definition}
In this work, we pre-train a graph encoder for PKG. The graph encoder preserves heterogeneous semantics of items, including both the features of items and their associated relations. 
The ultimate goal for pre-training a graph encoder is to tackle the ZSIR problem. 
Though the encoder $\text{Enc}(\cdot)$ is able to generate embeddings for all nodes in PKG $\mathcal{G}$, zero-shot items may not be present in the pre-training stage of the graph encoder. 
Thus, we require that the graph encoder can perform inductive on new items, defined below.
\begin{definition}
    \textbf{Inductive Inference.} Given a PKG $\mathcal{G}$ and a pre-trained PKG encoder $\text{Enc}(\cdot)$, suppose there is a zero-shot item $i^{*}$ with edges $\mathcal{E}_{i}^{*}$. The inductive inference task is to encode an updated graph, $\text{Enc}(\mathcal{G}^{*})$, where $\mathcal{G}^{*}$ is constructed by including $i^{*}$ and $\mathcal{E}_{i}^{*}$ into $\mathcal{G}$.
\end{definition}
After inductive inference, we have embeddings for all items, including both warm items and zero-shot items. Thus, we can resolve the ZSIR task.

\begin{figure*}[]
\centering
\includegraphics[width=0.85\textwidth]{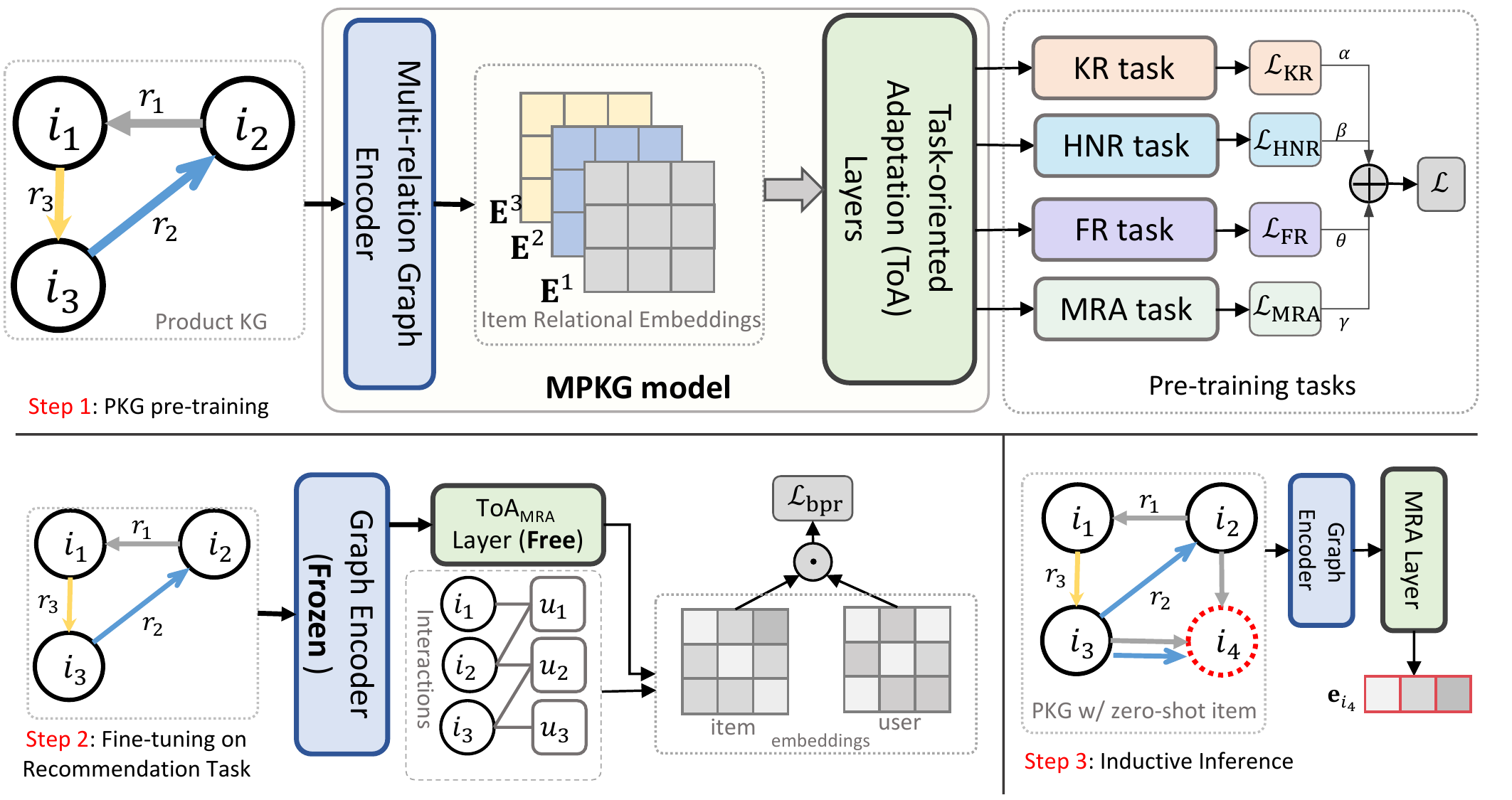}
\caption{The framework of our proposed method. In step 1, we pre-train the MPKG model by using multiple pre-training tasks upon the PKG. Each task is associated with one type of ToA layer.
Next in step 2, we fine-tune the model on the recommendation task with user-item interactions, which has frozen graph encoder and free ToA$_{\text{MRA}}$ layer parameters. 
Finally, in step 3 we conduct inductive inference of the zero-shot item $i_4$.  Best viewed in colors.}
\label{fig:model_architecture}
\end{figure*}
\section{Proposed Method}
In this section, we introduce the proposed framework \modelname for pre-training PKG encoder and fine-tuning the pre-trained model to the ZSIR task. The overall framework is shown in Figure~(\ref{fig:model_architecture}).

\subsection{Product Knowledge Graph Construction}
Constructing a PKG for pre-training requires two crucial factors: 
(1) the universal features from item generic information;  
(2) item-item connections derived from either meta-data or user-item interactions.
To be specific, the universal item features are task-invariant item generic features.
For example, 
in this paper, we extract items feature embeddings, $\mathbf{X}$, from the pre-trained BERT~\cite{devlin2018bert} by using the concatenated description and title texts of items as input.
We also analyze the effects of other PLMs~\cite{sanh2019distilbert,reimers2019sentence}.
Item-item connections are derived from the collected feedback.
Inspired by previous works~\cite{xu2020product,xu2020knowledge}, our PKG consists of multiple item relationships, including complement, co-view, substitute, etc, which are extracted from user interaction data. 

\subsection{Multi-Relation Graph Encoder}
In the pre-training stage, we first encode the PKG to obtain item embeddings over various relationships, which is shown in the upper left component in Fig.~(\ref{fig:model_architecture}).
The semantics of PKG contains multiple item-item relations. Therefore, during pre-training of the graph encoder, we simultaneously ingest both relations and node features for encoding the graph. 
We ensure the graph encoder has the inductive inference ability such that zero-shot item embeddings can be inferred during the evaluation stage. 

Motivated by the effectiveness of existing works~\cite{wei2022contrastive, wang2021self, he2020lightgcn}, we adopt a graph encoder based on the message-passing framework for each edge type. 
Specifically, for each edge type $r\in\mathcal{R}$, we extract the $r$-PKG as $\mathcal{G}^r=\{\mathcal{I}, \mathcal{E}^r, \mathbf{X}\}$.
Following SGCN~\cite{wu2019simplifying}, we adopt the encoder to obtain the item embeddings with $M$ layers of message aggregation as follows:
\begin{equation}
    \text{Enc}(\mathcal{G}^r) \rightarrow \mathbf{E}^{r} = \left(\mathbf{D}^{-\frac{1}{2}}\tilde{\mathbf{A}}\mathbf{D}^{-\frac{1}{2}}\right)^M\mathbf{X}\mathbf{W}^{r},
\end{equation}
where $\mathbf{E}^{r}\in \mathbb{R}^{|\mathcal{I}|\times d}$ is the embeddings of items w.r.t. relation $r$, $\mathbf{D}$ denotes the degree matrix of $\tilde{\mathbf{A}}$, $\tilde{\mathbf{A}}$ is the adjacency matrix with self-loop for $r$-PKG, and $\mathbf{W}^{r}\in \mathbb{R}^{d \times d}$ is the weight matrix.
The advantage of this simple form and the removal of activation in each aggregation layer allows the pre-computation of high-order neighborhood connectivity matrix, which significantly increases the efficiency.
Also, since the multi-layer message aggregation process, \textit{i.e.}, the term $\left(\mathbf{D}^{-\frac{1}{2}}\tilde{\mathbf{A}}\mathbf{D}^{-\frac{1}{2}}\right)^M$ can be decoupled from the feature transformation step, \textit{i.e.} the term $\mathbf{X}\mathbf{W}^{r}$, 
we can update the PKG with zero-shot items and conduct the message-passing directly on updated PKG, thus ensuring the inductive inference ability.

\subsection{Task-oriented Adaptation Layer}
Given $|\mathcal{R}|$ relationships, we obtain $|\mathcal{R}|$ item embedding matrices.
However, our final goal is to resolve the ZSIR task,
which requires the fusion of embeddings from all relations. 
Due to the domain discrepancy from our PKG pre-training tasks to the ZSIR task, 
we adapt the embeddings to different tasks such that semantics from multiple relations can be properly fused.  
Let $\{\mathbf{E}^{r}\}|_{r\in\mathcal{R}}$ 
denote the item embeddings for 
$|\mathcal{R}|$ relations. We define the fused embedding for a specific task $t$ as:
\begin{equation}
    \mathbf{E}_{t} = \text{ToA}_{t}(\{\mathbf{E}^{r}\}|_{r\in\mathcal{R}}),
\end{equation}
where $\text{ToA}_{t}$ can be arbitrary read-out functions, such as concat, mean-pooling, weighted-sum, etc. 
In the next sections, we discuss defining ToA layers for various pre-training tasks and fine-tuning on down-stream tasks.

\subsection{Multi-task Pre-training}
Our approach pre-trains a multi-relation graph encoder on four tasks for the PKG, which are presented as the step 1 in Figure~(\ref{fig:model_architecture}). 
In this section, we introduce these pre-training tasks and define the corresponding ToA layers.

\subsubsection{Knowledge Reconstruction (KR)}
Let $(i,r,j)$ denote a knowledge triplet, where items $i$ and $j$ are connected by relation $r$. To preserve the original semantics from each relation, we adopt a knowledge reconstruction task with respect to each relation.
To be concrete, we propose a link prediction task for each relation. In our link prediction task, the encoded item embeddings $\mathbf{E}^{r}$ must effectively reconstruct the item-item knowledge triplets under relation $r$.
Therefore, in this knowledge reconstruction task for relation $r$, the ToA layer uses only the embedding $\mathbf{E}^{r}$. 
We calculate the knowledge reconstruction score $s_{ij}^{r}$ as follows:
\begin{equation}
    s_{ij}^{r} = \sigma(\mathbf{E}^{r}_{i}\cdot\mathbf{E}^r_{j}),
\end{equation}
where $\sigma(\cdot)$ denotes the sigmoid activation function and $\mathbf{E}^{r}_{i}$ and $\mathbf{E}^{r}_{j}$ represent the embeddings under relation $r$ for item $i$ and $j$, respectively.
Hereafter, we develop the item knowledge link reconstruction loss, $\mathcal{L}_{\text{KR}}$, as the BCE loss between positive triplet and negative triplet and sum over all relations:
\begin{equation}
    \mathcal{L}_{\text{KR}} = \sum_{r\in \mathcal{R}}-\frac{1}{|\mathcal{E}^r|}\sum_{(i, j)\in\mathcal{E}^r} \left(\log s_{ij}^{r} + \log (1-s_{ij_{-}}^{r})\right),
\end{equation}
where $\mathcal{E}^r$ denotes all links under relation $r$ and ${(i, j_{-})\notin\mathcal{E}^r}$ is a negative sample to pair with the positive link.  

\subsubsection{High-order Neighbor Reconstruction (HNR)}
While the knowledge reconstruction task encourages the graph encoder to be relation-aware, due to sparsity of PKGs, it is insufficient to only consider direct neighbors.
Thus, we leverage the higher-order neighbors in the PKG to fully reconstruct the semantics. 
Specifically, we enhance the embeddings by reconstructing the $K$-order neighbors, regardless of relationships, which is defined as the High-order Neighbor Reconstruction~(HNR) task. 
This task simultaneously incorporates semantics from all relations. 
Hence, 
we define the ToA layer for this task as the concatenation for all embeddings,
\begin{equation}
    \mathbf{E}_{\text{HNR}} = \text{ToA}_{\text{HNR}}(\{\mathbf{E}^{r}\}|_{r\in\mathcal{R}}) = \text{Concat}(\{\mathbf{E}^{r}\}|_{r\in\mathcal{R}}),
\end{equation}
where $\mathbf{E}_{\text{HNR}}$ and $\text{ToA}_{\text{HNR}}$ denotes the item embeddings and ToA layer for this HNR task respectively.
We first collect the $K$-order neighbors of each item, denoted as $\mathcal{N}_K(i)$. 
Then the neighbor reconstruction score $a_{ij}$ between item $i$ and $j$ is defined as the soft dot-product:
\begin{equation}
    a_{ij} = \sigma(\mathbf{E}_{\text{HNR}}({i}) \cdot \mathbf{E}_{\text{HNR}}({j})), 
\end{equation}
where $\mathbf{E}_{\text{HNR}}({i})$ and $\mathbf{E}_{\text{HNR}}({j})$ represent the HNR embedding for item $i$ and $j$, respectively.
Finally, we optimize the task with BCE loss as follows:
\begin{equation}
    \mathcal{L}_{HNR} = -\sum_{i\in \mathcal{I}}\sum_{j\in \mathcal{N}_K(i)} \left(\log a_{ij} + \log(1- a_{ij_{-}})\right),
\end{equation}
where $j$ denote a K-hop neighbor of item $i$ and  $j_{-}$ denotes a negative sample to pair with $j$ such that $j_{-}\in \mathcal{I}\setminus \mathcal{N}_K(i)$

\subsubsection{Feature Reconstruction (FR)}
The universal item features encode the basic item generic information and benefit the inductive inference for zero-shot items. 
However, since universal item features are extracted from PLMs, there is a large semantic divergence between the universal item features and output from multi-relation graph encoder.  
Therefore, we propose to use the feature reconstruction~(FR) task to optimize the graph encoder such that semantic divergence is mitigated.
Concretely, we propose to use the item embeddings from graph encoder to reconstruct the universal item features via a decoder. 
For this task, semantics from all relations are 
 also harnessed. 
Hence, we define the ToA layer to be the same as in HNR task, \textit{i.e.} the concatenation, as follows:
\begin{equation}
    \mathbf{E}_{\text{FR}} = \text{ToA}_{\text{FR}}(\{\mathbf{E}^{r}\}|_{r\in\mathcal{R}}) = \text{Concat}(\{\mathbf{E}^{r}\}|_{r\in\mathcal{R}}),
\end{equation}
where $\mathbf{E}_{\text{FR}}$ denotes the item embeddings for this FR task.
Then, we input this $\mathbf{E}_{\text{FR}}$ to a decoder Dec$(\cdot)$ such that the universal feature from PLMs can be reconstructed, formulated as follows:
\begin{equation}
    \tilde{\mathbf{X}} = \text{Dec}(\mathbf{E}_{\text{FR}}),
\end{equation}
where $\tilde{\mathbf{X}}$ is the feature decoded from the concatenated relational embeddings.
Though a wide range of decoders can tackle this FR task, we adopt one fully-connected layer as the decoder here in this paper. 
The reason is that a light-weight decoder is less complex to optimize and the output embeddings from graph encoder can be linearly aligned with universal features.
We leave the investigation of other types of decoders as future work.

Finally, we optimize this task under the measurement of $L_{2}$ losses between orignal features and reconstructed features as follows:
\begin{equation}
    \mathcal{L}_{\text{FR}} = \sum_{i\in \mathcal{I}}\|\mathbf{X}_{i} - \tilde{\mathbf{X}}_{i}\|_{2}^{2},
\end{equation}
where $\mathbf{X}_{i}$ and $\tilde{\mathbf{X}}_{i}$ are the universal and reconstructed features for item $i$, respectively. 

\subsubsection{Meta Relation Adaptation (MRA)}
Recall that the objective of pre-training a graph encoder is to yield embeddings for the items in the downstream ZSIR task. 
Nevertheless, due to the domain discrepency between PKG semantics and the ZSIR task, different item-item relations have unequal contributions. 
Therefore, we should devise a proper strategy to adapt relational embeddings to various tasks.
Since during the pre-training stage, we have no access to downstream data, we propose a novel Meta Relation Adaptation~(MRA) task.
To be concrete, we treat one relation $r$ as the target relation, and use embeddings from other relations to reconstruct the edges in $r$-PKG $\mathcal{G}^{r}$. We define this as the $r$-MRA task. 
Firstly, the ToA layer for $r$-MRA task is a weighted sum of all relational embeddings except the relation $r$ embeddings, which is formulated as:
\begin{equation}\label{eq:r-mra}
    \mathbf{E}_{r\text{-MRA}} = \text{ToA}_{r\text{-MRA}}(\{\mathbf{E}^{r}\}|_{r\in\mathcal{R}}) = {\sum_{r\in{\mathcal{R}_{-r}}} w_{r}}\mathbf{E_r},
\end{equation}
where $\mathcal{R}_{-r}$ denotes all relations but relation $r$, and  $w_{r} \in \mathbb{R}$ is a scalar weights, denoting the contrition of each relation embeddings in $\mathcal{R}_{-r}$. 
In this paper, we use a self-excitation layer~\cite{hu2018squeeze} to compute the weight $w_{r}$, which ingests the associated relation embeddings into two fully-connected layers and normalizes those weights w.r.t. each relation with a softmax function. 
The reason is self-excitation layer is easy to implement and fine-tune for new downstream tasks.
We leave other methods for calculating the weights in future works.
Next, we predict edges in $r$-PKG by a soft dot-product upon the $r$-MRA embedding. The prediction score $b_{ij}$ between item $i$ and $j$ is formulated as follows:
\begin{equation}
    b_{ij} = \sigma(\mathbf{E}_{r\text{-MRA}}(i) \cdot \mathbf{E}_{r\text{-MRA}}(j)),
\end{equation}
where $\mathbf{E}_{r\text{-MRA}}(i)$ and $\mathbf{E}_{r\text{-MRA}}(j)$ represent the $r$-MRA embeddings for items $i$ and $i$, respectively.
The intuition for this meta relation adaption task is to simulate the process of adapting relation embeddings to new tasks.
The $r$-MRA task views the edge prediction task on $r$ relation as a new task and train the encoder to adapt the embeddings from other relation sematics to relation $r$. 
In this way, the encoder would have more generalizatio ability and endows the $\text{ToA}_{r\text{-MRA}}(\cdot)$ layer more flexibility for downstream task adaptation, thus resolving the domain discrepency problem betwen PKG semantics and ZSIR task. 
We will introduce how to fine-tune this layer in ZSIR task in the next section.

Next, we optimize the MRA tasks for all relations via MSE loss as follows:
\begin{equation}
    \mathcal{L}_{\text{MRA}} = \sum_{r\in\mathcal{R}} -\frac{1}{|\mathcal{E}^{r}|}\sum_{(i, j)\in\mathcal{E}^r} \left(\log b_{ij} +  \log (1-b_{ij_{-}})\right),
\end{equation}
where $\mathcal{E}^r$ denotes all edges under relation $r$ and ${(i, j_{-})\notin\mathcal{E}^r}$ is a negative sample to pair with the positive edge. 

\subsubsection{Final Pre-Training Loss}
We present the entire training framework as a multi-task training framework.
The final loss is calculated as the weighted sum of four proposed losses:
\begin{equation}
\label{eq:final_loss}
    \mathcal{L} = \alpha\mathcal{L}_{\text{KR}} + \beta\mathcal{L}_{\text{FR}} + \theta\mathcal{L}_{\text{HNR}} + \gamma\mathcal{L}_{\text{MRA}},
\end{equation}
where $\alpha$, $\beta$, $\theta$, and $\gamma$ are hyper-parameters, and we choose them based on the best performance on the validation set. 

\subsection{Model Fine-tuning}
In general, we could fine-tune the proposed model on any new tasks. 
We could update parameters in the graph encoder $\text{Enc}(\cdot)$ and ToA layers by defining new objective functions for new tasks.
This work mainly fine-tunes the ToA$_{\text{MRA}}$ layers for all relations as it is most relevant to the ZSIR task and more efficient to adapt without loading the entire PKG again in ZSIR.
Hence, we only discuss how to fine-tune ToA$_{\text{MRA}}$ layers on ZSIR task in this paper.
An example process is given in the step 2 of  Figure~(\ref{fig:model_architecture}).
Any other tasks can be investigated in analogy. 

Concretely, 
we update ToA$_{\text{MRA}}$ layers with a recommendation objective function.
Given the user-item interaction data, denoted as $\mathcal{D} = \{{(u,i)| u \in \mathcal{U}, i \in \mathcal{I}}\}$ where $\mathcal{U}$ is the user set, 
we optimize the pre-trained \modelname with only ToA$_{\text{MRA}}$ layers as free parameters and all other parameters are fixed.
During the fine-tuning stage, the item embeddings are produced similar to Eq.~(\ref{eq:r-mra}), but involving all relations in PKG, denoted as $\mathbf{E}_{\text{MRA}} = {\sum_{r\in{\mathcal{R}_{r}}} w_{r}}\mathbf{E_r}$. 
Recall that $w_r$ represents the contribution of each relation $r$ and computed via self-excitation over the relational embedding $\mathbf{E}_{r}$. 
Hence, we optimize the self-excitation layers such that the contribution of each relation towards recommendation task can be characterized. 

The recommendation task is to predict ranking scores between items and users. 
For each pair $(u,i)$, we represent the user representation $\mathbf{e}_{u}$ as the mean aggregation for all interacted items, formulated as $\mathbf{e}_{u} = \frac{1}{|\mathcal{D}_{u}|}\sum_{i \in \mathcal{D}_{u}}\mathbf{E}_{\text{MRA}}(i)$, where $\mathcal{D}_{u}$ is the interacted items for user $u$ and $\mathbf{E}_{\text{MRA}}(i)$ denotes the output embedding for item $i$. 
Then, we calculate the ranking score $p_{ui}$ between user $u$ and item $i$ via the dot-product similarity as follows:
\begin{equation}\label{eq:ranking_score}
    p_{ui} = \mathbf{e}_{u} \cdot \mathbf{E}_{\text{MRA}}(i).
\end{equation}
Finally, we fine-tune the model via the BPR loss~\cite{rendle2012bpr} as follows:
\begin{equation}
    \mathcal{L}_{bpr} = \sum_{(u,i)\in \mathcal{D}}-\log \sigma\left(p_{ui} - p_{ui_{-}}\right),
\end{equation}
where $i_{-}$ is a negative item such that $(u,i_{-}) \notin \mathcal{D}$ for user $u$.
After optimization, ToA$_{\text{MRA}}$ layers are adapted to the recommendation task.  
Note that we could use any other functions to produce the final representation of users and items for recommendation task.
This paper investigates the above methods as it is the minimal way to verify the effectiveness of the \modelname framework, no additional parameters being introduced during fine-tuning stage. 

Hereafter, we utilize the fine-tuned model to conduct inductive inference for the zero-shot items, which is demonstrated in the step 3 of Figure~\ref{fig:model_architecture}.
Finally, the prediction scores between users and all items are calculated as in Eq.~(\ref{eq:ranking_score}.)

\section{Experiments}
In this section, we demonstrate the effectiveness of our proposed universal pre-training PKG framework in several perspectives.
We answer the following Research Questions~(RQs) to validate the superiority:
\begin{itemize}[leftmargin=*]
    \item \textbf{RQ1: }Does \modelname generalize to downstream ZSIR tasks, especially for zero-shot items?
    \item \textbf{RQ2: }Does \modelname yield better universal item embeddings than other state-of-the-art models?
    \item \textbf{RQ3: }What are the contributions of multiple pre-training tasks for  \modelname?
    \item \textbf{RQ4: }Does using different variants of base models affect the pre-training?
\end{itemize}

\subsection{Data Preparation}
We conduct the experiments on the largest category \textit{Home and Kitchen} category in Xmarket dataset\footnote{\url{https://xmrec.github.io/}}.
The dataset consists of 18 markets, of which each has user-item reviews and item-item relationships as meta-data. We utilize the item-item relationships in meta-data as the PKG pre-training item relationships, including \textit{alsoViewed, alsoBought, boughtTogether} as these are widely used item relationships for recommendation~\cite{zhao2017improving, yan2022personalized, zhou2022decoupled}. We aggregate item-item relationships pairs from all markets and construct the PKG.
We list statistics of user-item interaction data of all markets in Table~\ref{tab:interaction_dataset}. The data statistics of the product knowledge graph are in Table~\ref{tab:pkg_stats}. 
We concatenate the description and title texts as the universal textual information, and we extract the item universal features $\mathbf{X}$ using a pre-trained language model.


We rank the user-item interactions in chronological order. We use data in the earliest 80\% time for training, the following 10\% time for validation, and the last 10\% period for testing. 
The items appearing in the training data are the train item set. For validation and testing items appearing in the train item set, we denote them as warm items, otherwise, we denote them as zero-shot (zs) items. 
To avoid the data leakage problem
we delete all the cold items from PKG during training.

\begin{table*}[]
\centering
\caption{Home and Kitchen Dataset User-Item Interactions Statistics. For each market, 
 statistics are reported in the following format: \#of users\slash \#of items\slash \#of user-item interactions. 
 We filter out markets with less than 1,000 users. }
\label{tab:interaction_dataset}
\begin{tabular}{|c|c|c|c|c|c|}
\hline
\textbf{Brazil(br)} & \textbf{Japan(jp)} & \textbf{Mexico(mx)} & \textbf{Italian(it)} & \textbf{France(fr)} & \textbf{Spain(es)} \\ \hline
1.7K/60K/10.5K & 1.9K/5.6K/14K & 3.5K/7K/16K & 3.7K/11K/20K & 5.7K/16K/38K & 5.9K/10K/29K \\ \hline
\textbf{Australia(au)} & \textbf{Germany(de)} & \textbf{India(in)} & \textbf{Canada(ca)} & \textbf{United States(us)} & \textbf{United Kingdom(uk)} \\ \hline
13K/27.6K/121K & 18K/31K/122K & 22K/20K/114K & 30K/38K/208K & 1.7K/8K/8K & 251K/66K/1.8M \\ \hline
\end{tabular}
\end{table*}


\begin{table*}[]
\centering
\caption{Product Knowledge Graph Statistics.}
\label{tab:pkg_stats}
\begin{tabular}{|c|c|c|c|c|c|}
\hline
Edge Type & alsoBought & alsoViewed & boughtTogether & Total \\ \hline
\#of items/\#of edges & \multicolumn{1}{r|}{93,659/2,044,418} & \multicolumn{1}{r|}{86,481/1,048,779} & \multicolumn{1}{r|}{64,574/115,386} & \multicolumn{1}{r|}
{97,626/3,208,583} \\ \hline
\end{tabular}
\end{table*}

\subsection{Evaluation Tasks}
We present the effectiveness of our proposed pre-training PKG framework via two evaluation tasks, \textit{i.e.} the \textit{knowledge prediction} task and \textit{zero-shot item-based recommendation}~(ZSIR) task.
The knowledge prediction task assesses the ability of our pre-trained GNN in recovering the semantics between items in the PKG. Specifically, the knowledge prediction task predicts the knowledge triplet links associated with items as head entities.
The ZSIR task assesses the inference ability of \modelname on a downstream task.

The performance of both tasks is evaluated on all items and zero-shot items settings.
For all downstream tasks, we generate the top-N ranking list from either the all item candidates, or only the test zero-shot items. We report the overall performance on both settings to demonstrate the ability of our model in pre-training universal item embeddings. The inductive inference introduced in Section 3 infers the embeddings of zero-shot items with the complete PKG in test time. We adopt the standard ranking performance metrics Recall@N and Mean Reciprocal Rank~(MRR) as evaluation metrics.
We report the testing performance based on the grid-searched best validation performance.

\subsection{Baselines and Implementation}
To validate the effectiveness of the proposed framework, we compare the model with the following two groups of related baselines: (1) Triplet-based heterogeneous graph methods, including TransE~\cite{bordes2013translating}, TransD~\cite{ji2015knowledge}, DistMult~\cite{zhang2018knowledge}, and TransH~\cite{wang2014knowledge}; (2) Heterogeneous graph models, including GPT-GNN~\cite{hu2020gpt} with a generative graph model framework and HeCo~\cite{wang2021self} with the self-supervised graph learning architecture. 

We implement \modelname in PyTorch and conduct the experiments with 4 V100 GPUs. We grid search important hyper-parameters in baselines and the proposed \modelname. During the pre-training stage, we can only access the knowledge triplets and we select the best pre-training \modelname based on the validation performance on validation set of knowledge triplets predictions. For all methods, we search the hidden dimension from $\{64, 128\}$, the L2 regularization weight from $\{1e-3, 1e-2, 1e-1, 5e-1\}$, the learning rate from $\{1e-3, 1e-4, 5e-3\}$, the batch size is set to be 256, the base GNN is SGC~\cite{wu2019simplifying}, and the number of GNN layers is default at 3. For all triplet-based heterogeneous graph baselines, we search the hidden dimension and L2 regularization weight. For heterogeneous graph model GPT-GNN~\cite{hu2020gpt}, we additionally search its attribute generation loss ratio from $\{0.1, 0.3, 0.5, 0.7, 0.9\}$ and the queue size from $\{128, 256, 512\}$. For HeCo~\cite{wang2021self}, we further search its dropout rate for features and attentions from $\{0.1, 0.3, 0.5, 0.7, 0.9\}$.

\begin{table*}[]
\centering
\caption{ZSIR Task Results on All Items Comparison. The best models are bolded and the second-best are underlined.}
\label{tab:itembasedrec_all}
\resizebox{\textwidth}{!}{%
\begin{tabular}{c|ccccccc|ccccccc}
\hline
 & \multicolumn{7}{c|}{NDCG@20} & \multicolumn{7}{c}{MRR} \\ \hline
Model & br & mx & es & in & ca & uk & us & br & mx & es & in & ca & uk & us \\ \hline
TransE & 0.0357 & 0.0250 & 0.0196 & 0.0153 & 0.0149 & 0.0126 & 0.0166 & 0.0298 & 0.0186 & 0.0163 & 0.0121 & 0.0100 & 0.0080 & 0.0102 \\
TransD & 0.0345 & 0.0251 & 0.0189 & 0.0157 & 0.0148 & 0.0127 & 0.0159 & 0.0275 & 0.0199 & 0.0153 & 0.0117 & 0.0103 & 0.0080 & 0.0103 \\
TransH & 0.0393 & 0.0289 & 0.0208 & 0.0164 & 0.0160 & 0.0129 & 0.0197 & 0.0317 & 0.0229 & 0.0187 & 0.0143 & 0.0107 & 0.0083 & 0.0126 \\
DistMult & 0.0344 & 0.0233 & 0.0202 & 0.0159 & 0.0168 & 0.0125 & 0.0155 & 0.0253 & 0.0207 & 0.0195 & 0.0129 & 0.0111 & 0.0075 & 0.0088 \\
GPT-GNN & {\ul 0.0397} & {\ul 0.0304} & {\ul 0.0222} & {\ul 0.0182} & 0.0174 & {\ul 0.0140} & 0.0210 & {\ul 0.0329} & 0.0242 & {\ul 0.0251} & {\ul 0.0169} & 0.0113 & {\ul 0.0096} & {\ul 0.0158} \\
HeCo & 0.0384 & 0.0285 & {\ul 0.0222} & 0.0175 & {\ul 0.0175} & 0.0132 & {\ul 0.0217} & 0.0323 & {\ul 0.0246} & 0.0234 & 0.0156 & {\ul 0.0125} & 0.0090 & 0.0157 \\
\modelname & \textbf{0.0542} & \textbf{0.0433} & \textbf{0.0407} & \textbf{0.0243} & \textbf{0.0204} & \textbf{0.0162} & \textbf{0.0257} & \textbf{0.0413} & \textbf{0.0332} & \textbf{0.0346} & \textbf{0.0208} & \textbf{0.0166} & \textbf{0.0139} & \textbf{0.0208} \\ \hline
Impro. & 36.52\% & 42.43\% & 83.33\% & 33.52\% & 16.57\% & 15.71\% & 18.43\% & 25.53\% & 34.96\% & 37.85\% & 23.08\% & 32.8\% & 44.79\% & 31.65\%
 \\ \hline
\end{tabular}%
}

\end{table*}

\begin{table*}[]
\centering
\caption{ZSIR Task Results on Zero-Shot Items Comparison. The second-best and the best models are underlined and bolded.}
\label{tab:itembasedrec_cold}
\resizebox{\textwidth}{!}{%
\begin{tabular}{c|ccccccc|ccccccc}
\hline
 & \multicolumn{7}{c|}{NDCG@20} & \multicolumn{7}{c}{MRR} \\ \hline
Model & br & mx & es & in & ca & uk & us & br & mx & es & in & ca & uk & us \\ \hline
TransE & 0.0467 & 0.0365 & 0.0312 & 0.0206 & 0.0162 & 0.0130 & 0.0215 & 0.0333 & 0.0245 & 0.0216 & 0.0150 & 0.0115 & 0.0096 & 0.0143 \\
TransD & 0.0454 & 0.0332 & 0.0304 & 0.0213 & 0.0175 & 0.0132 & 0.0205 & 0.0313 & 0.0227 & 0.0228 & 0.0130 & 0.0121 & 0.0088 & 0.0144 \\
TransH & 0.0501 & 0.0404 & 0.0312 & 0.0240 & 0.0197 & 0.0152 & 0.0252 & 0.0364 & 0.0280 & 0.0202 & 0.0167 & 0.0114 & 0.0095 & 0.0181 \\
DistMult & 0.0452 & 0.0332 & 0.0309 & 0.0202 & 0.0174 & 0.0129 & 0.0198 & 0.0290 & 0.0241 & 0.0227 & 0.0151 & 0.0125 & 0.0086 & 0.0126 \\
GPT-GNN & {\ul 0.0528} & {\ul 0.0426} & {\ul 0.0319} & 0.0236 & {\ul 0.0194} & {\ul 0.0147} & {\ul 0.0342} & {\ul 0.0404} & {\ul 0.0302} & {\ul 0.0229} & 0.0178 & {\ul 0.0124} & {\ul 0.0109} & 0.0240 \\
HeCo & 0.0508 & 0.0403 & 0.0326 & {\ul 0.0240} & 0.0210 & 0.0160 & 0.0324 & 0.0401 & 0.0288 & 0.0219 & {\ul 0.0180} & 0.0117 & 0.0101 & {\ul 0.0245} \\
\modelname & \textbf{0.0601} & \textbf{0.0481} & \textbf{0.0452} & \textbf{0.0270} & \textbf{0.0227} & \textbf{0.0181} & \textbf{0.0358} & \textbf{0.0431} & \textbf{0.0348} & \textbf{0.0358} & \textbf{0.0213} & \textbf{0.0168} & \textbf{0.0129} & \textbf{0.0262} \\ \hline
Impro. & 13.83\% &	12.91\% &	41.69\% &	12.5\% &	17.01\% &	23.13\% &	4.68\% &	6.68\% &	15.23\% &	56.33\% &	18.33\% &	35.48\% &	18.35\% &	6.94\% \\
 \hline
\end{tabular}%
}
\end{table*}

\subsection{ZSIR Task Performance~(RQ1)}

We conduct the ZSIR evaluation in multiple markets. 
We report the performance on all items recommendation in Table~\ref{tab:itembasedrec_all} and the performance on only zero-shot items recommendation in Table~\ref{tab:itembasedrec_cold}. We only list 7 markets due to the space limitation. The first three from the left are the smallest 3 markets while the remaining 4 markets are the largest 4 markets.
From both tables, we have the following observations:
\begin{itemize}[leftmargin=*]
    \item In both all items and zero-shot items recommendations, our proposed \modelname consistently achieves the best performance in all markets and all metrics. The relative improvements range from 23.08\% to 83.33\% in all items recommendation. For zero-shot items recommendation, the improvements are from 4.68\% to 56.33\%. These improvements demonstrate that the proposed \modelname framework successfully addresses the domain discrepancy between the PKG and the downstream ZSIR task in the zero-shot setting, which assumes no data is seen in the pre-training stage. We argue that the improvements come from the superior pre-training capability on handling multi-type item relationships and the adaptation layer to improve the generalization capability.
    \item The pre-training heterogeneous GNN baselines outperform the triplet-based methods. However, there is not a consistent winner among heterogeneous GNN baselines. This again demonstrates the importance of multi-type relations modeling in GNN.
    \item The improvements on low-resource markets are larger than the rich markets. For example, in all items recommendation, the low-resource markets have at least 36.53\% relative improvements in NDCG@20 while the larger markets have at most 33.52\%. This demonstrates that \modelname can benefit low-resource markets more than rich markets, indicating better generalization capability.
\end{itemize}



\subsection{Knowledge Prediction Comparison~(RQ2)}
In this section, we validate the pre-training effectiveness of the proposed \modelname in learning item-item relationships predictions, in both warm items~(seen items in training portion) and zero-shot items~(unseen items). The knowledge prediction task validates the capability of pre-training with product knowledge graph information over existing methods. 

\begin{figure}[]
\centering
     \begin{subfigure}[b]{0.23\textwidth}
         \centering
         \includegraphics[width=1\textwidth]{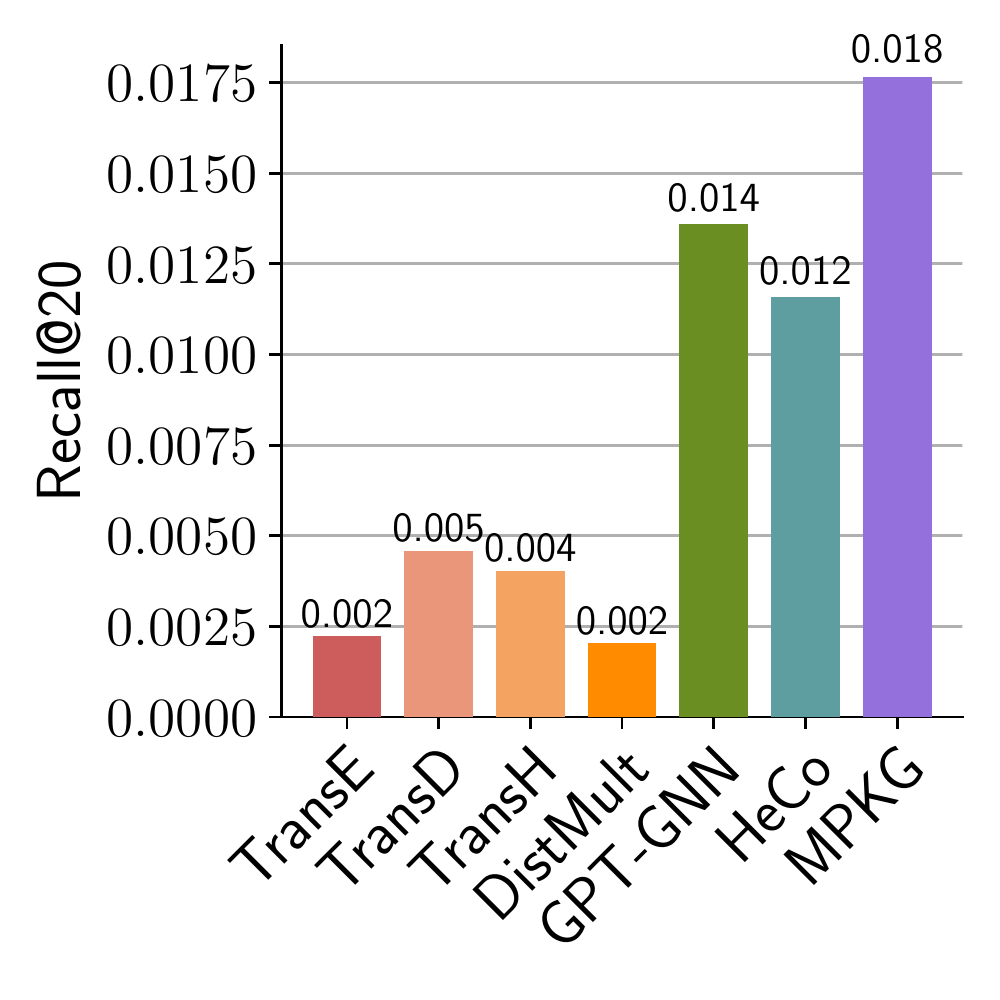}
         \caption{Comparison on Recall@20}
         \label{fig:kglink_pred_warm_recall}
     \end{subfigure}\hfill
     \begin{subfigure}[b]{0.23\textwidth}
         \centering
         \includegraphics[width=1\textwidth]{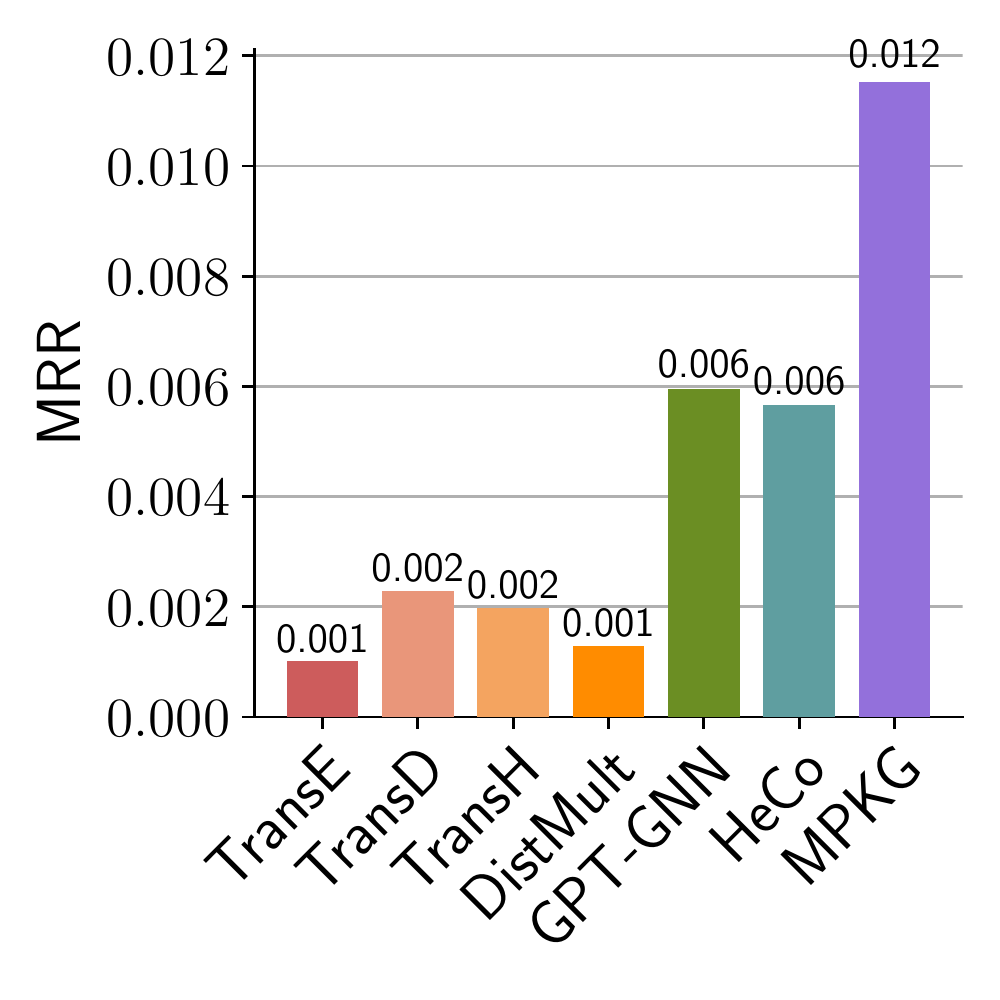}
         \caption{Comparison on MRR}
         \label{fig:kglink_pred_warm_mrr}
     \end{subfigure}
\caption{Knowledge Prediction on Warm (Seen) Items.}
\label{fig:kp_warm}
\end{figure}

\begin{figure}[]
\centering
     \begin{subfigure}[b]{0.23\textwidth}
         \centering
         \includegraphics[width=1\textwidth]{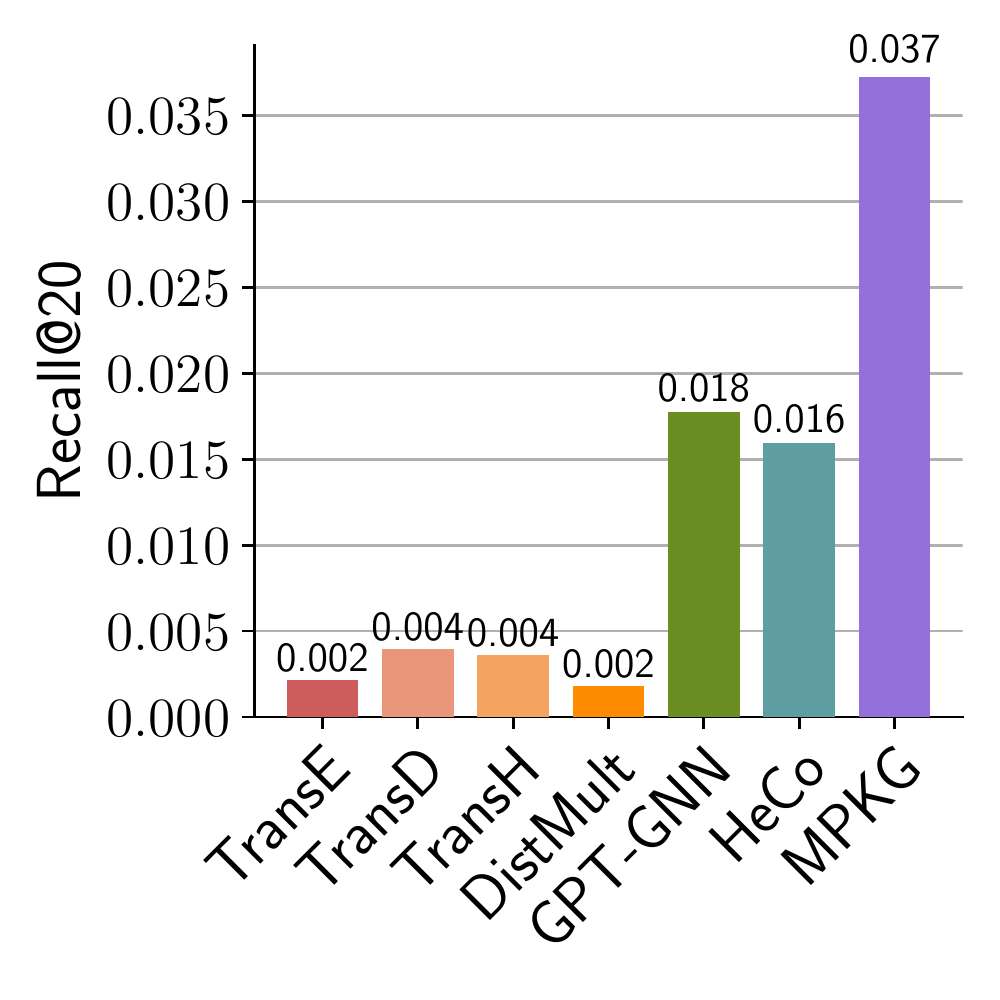}
         \caption{Comparison on Recall@20}
         \label{fig:kglink_pred_cold_recall}
     \end{subfigure}\hfill
     \begin{subfigure}[b]{0.23\textwidth}
         \centering
         \includegraphics[width=1\textwidth]{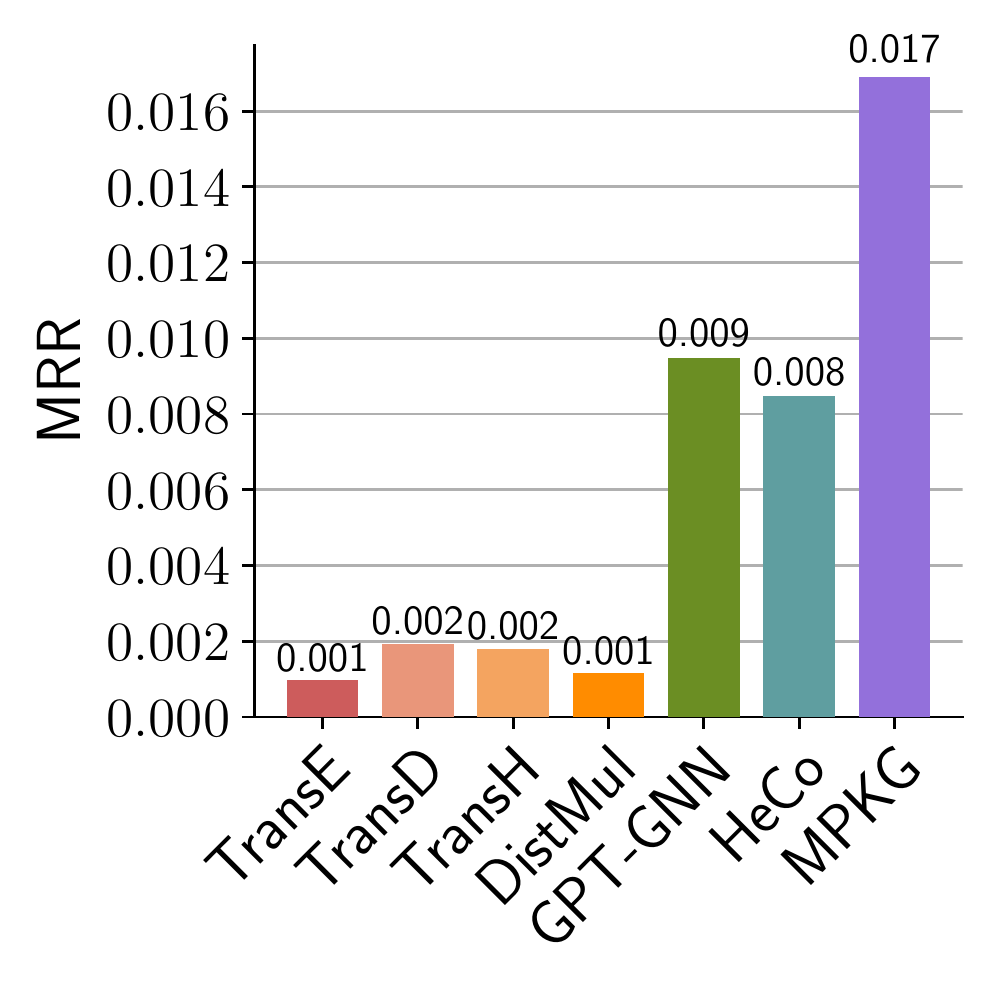}
         \caption{Comparison on MRR}
         \label{fig:kglink_pred_cold_mrr}
     \end{subfigure}
\caption{Knowledge Prediction on Zero-Shot Items.}
\label{fig:kp_cold}
\end{figure}

\subsubsection{Warm Items Comparison}
\label{sec:warm_kgpred}
The knowledge prediction performance of product knowledge graph triplets on warm items are shown in Figure~\ref{fig:kp_warm}. We report the Recall@20 and MRR in Figure~\ref{fig:kglink_pred_warm_recall} and Figure~\ref{fig:kglink_pred_warm_mrr}, respectively. We obtain the following observations from these comparisons:
\begin{itemize}[leftmargin=*]
    \item The proposed \modelname achieves the best warm item knowledge prediction performance in both metrics, with relative improvements from 28\% to 100\% in all metrics. We attribute this superior capability to the design of several proposed pre-training tasks as it mitigates the semantic divergence between generic information and item multi-relations.
    \item Among compared baselines, we observe that pre-training methods based on heterogeneous GNN (GPT-GNN, HeCo, and our \modelname) achieve better performances than triple-based methods. The heterogeneous GNN methods outperform triplet-based methods due to the stronger modeling capability of multi-relations in PKG while triplet-based methods only model direct connections and item features. 
\end{itemize}

\subsubsection{Zero-Shot Items Comparison}
We further conduct the knowledge prediction task on zero-shot items. 
The zero-shot item embedding inference is corresponding to the inductive inference as in the step 3 in Figure~\ref{fig:model_architecture} but without the fine-tuning step.
The performance is shown in Figure~\ref{fig:kp_cold}. We also report the Recall@20 and MRR in Fig.~(\ref{fig:kglink_pred_cold_recall}) and Fig.~(\ref{fig:kglink_pred_cold_mrr}), respectively. 
Zero-shot items evaluation verifies the induction capability of models and demonstrates the extent to which item embeddings generation can extend to zero-shot items. From the comparison, we have several observations:
\begin{itemize}[leftmargin=*]
    \item \modelname still achieves the best zero-shot item knowledge prediction performances in all metrics, with improvements from 88.9\% to 105.6\% over the best baseline model. 
    The superiority in knowledge prediction performances demonstrates the effectiveness of \modelname in generalizing to zero-shot items.
    \item Among the two categories of baselines approaches, pre-training methods based on heterogeneous GNN still achieve more satisfactory item embeddings learning than triplet-based methods. It further demonstrates the necessity of GNN in generalizing item embeddings learning.
\end{itemize}

\begin{figure}[]
\centering
     \begin{subfigure}[b]{0.2\textwidth}
         \centering
         \includegraphics[width=1\textwidth]{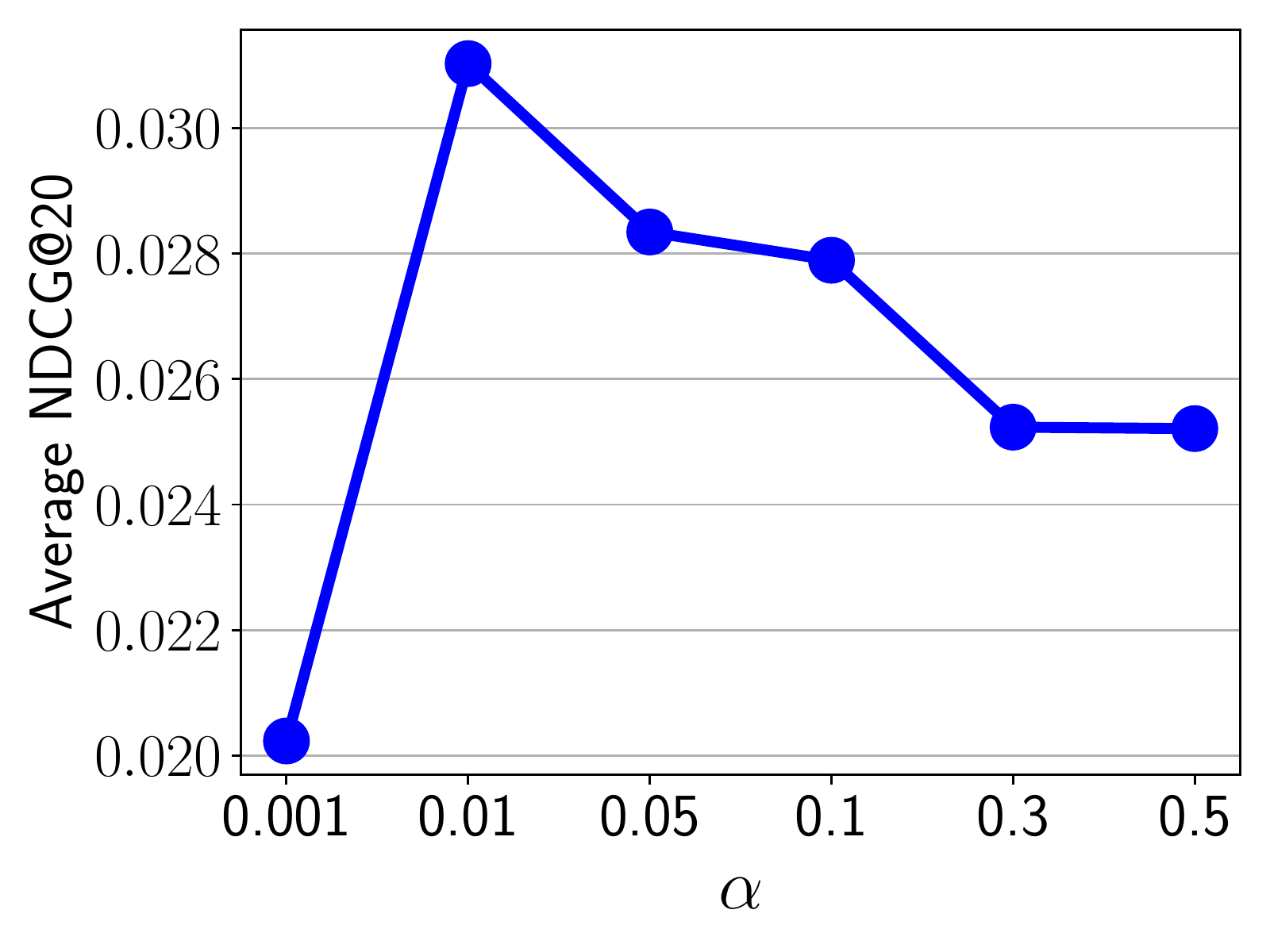}
         \caption{$\alpha$: weight of $\mathcal{L}_{\text{KR}}$}
         \label{fig:alpha_sens}
     \end{subfigure}\hfill
     \begin{subfigure}[b]{0.2\textwidth}
         \centering
         \includegraphics[width=1\textwidth]{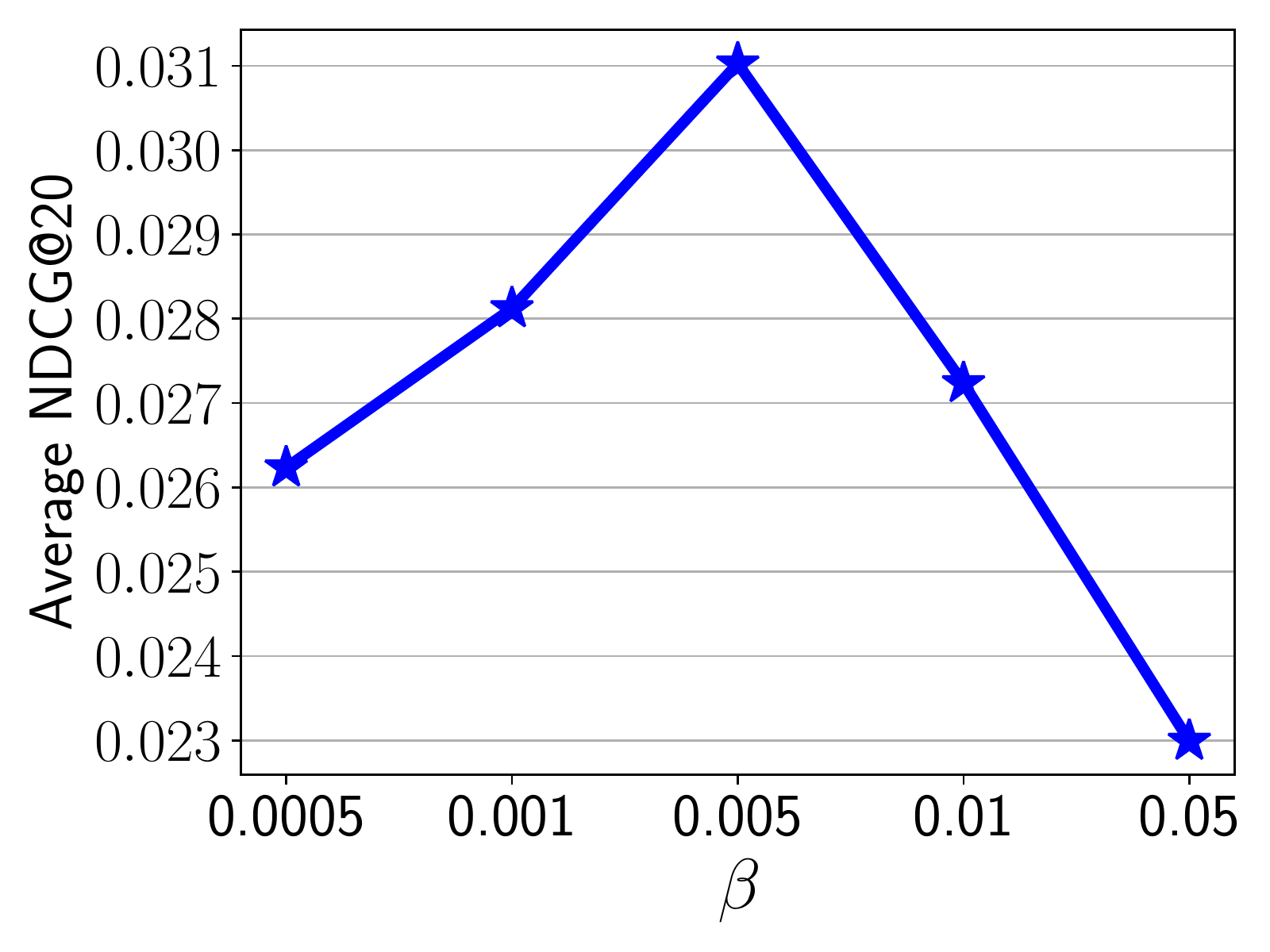}
         \caption{$\beta$: weight of $\mathcal{L}_{\text{FR}}$}
         \label{fig:beta_sens}
     \end{subfigure}\\
     \begin{subfigure}[b]{0.2\textwidth}
         \centering
         \includegraphics[width=1\textwidth]{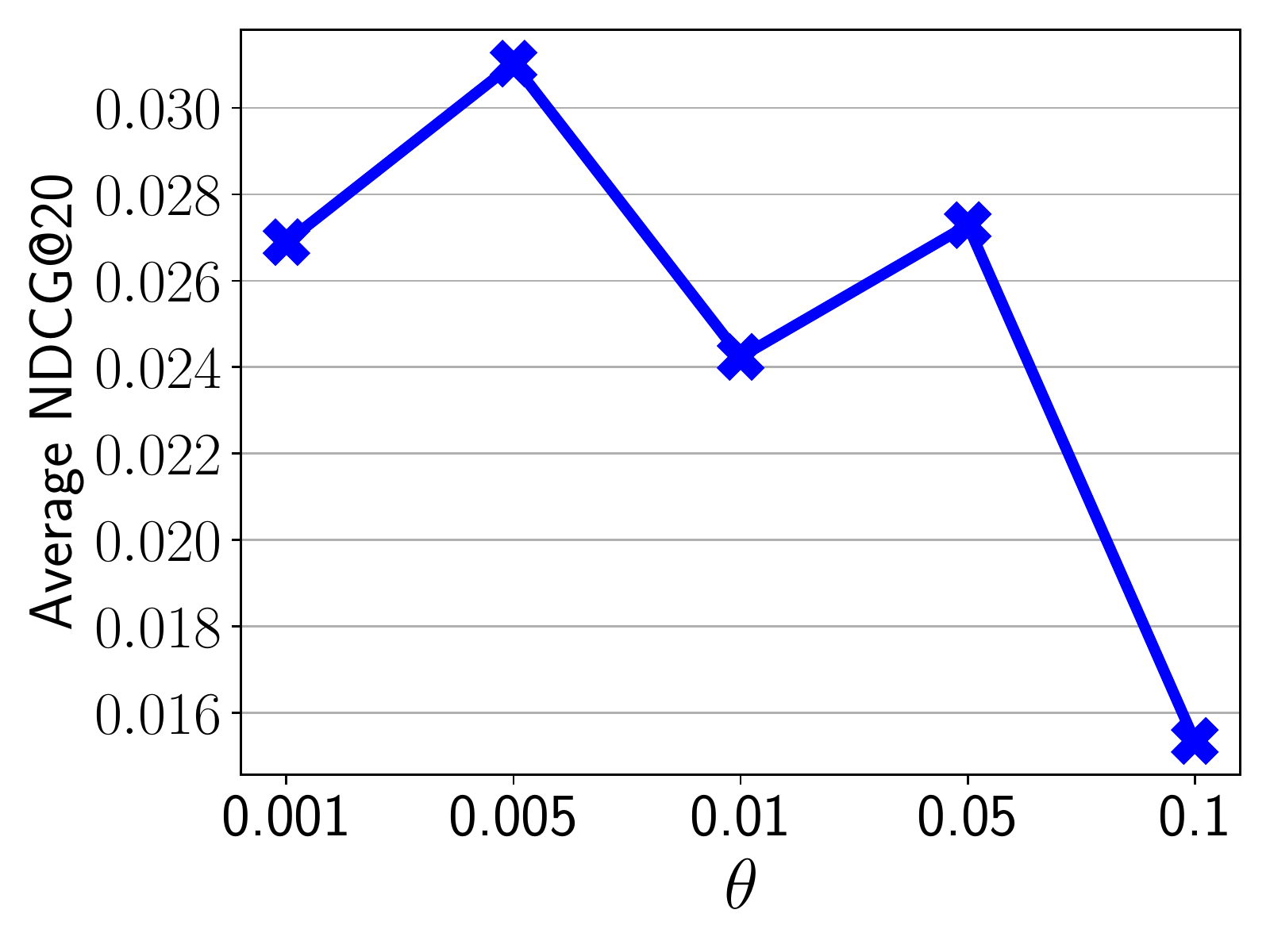}
         \caption{$\theta$: weight of $\mathcal{L}_{\text{HNR}}$}
         \label{fig:theta_sens}
     \end{subfigure}\hfill
     \begin{subfigure}[b]{0.2\textwidth}
         \centering
         \includegraphics[width=1\textwidth]{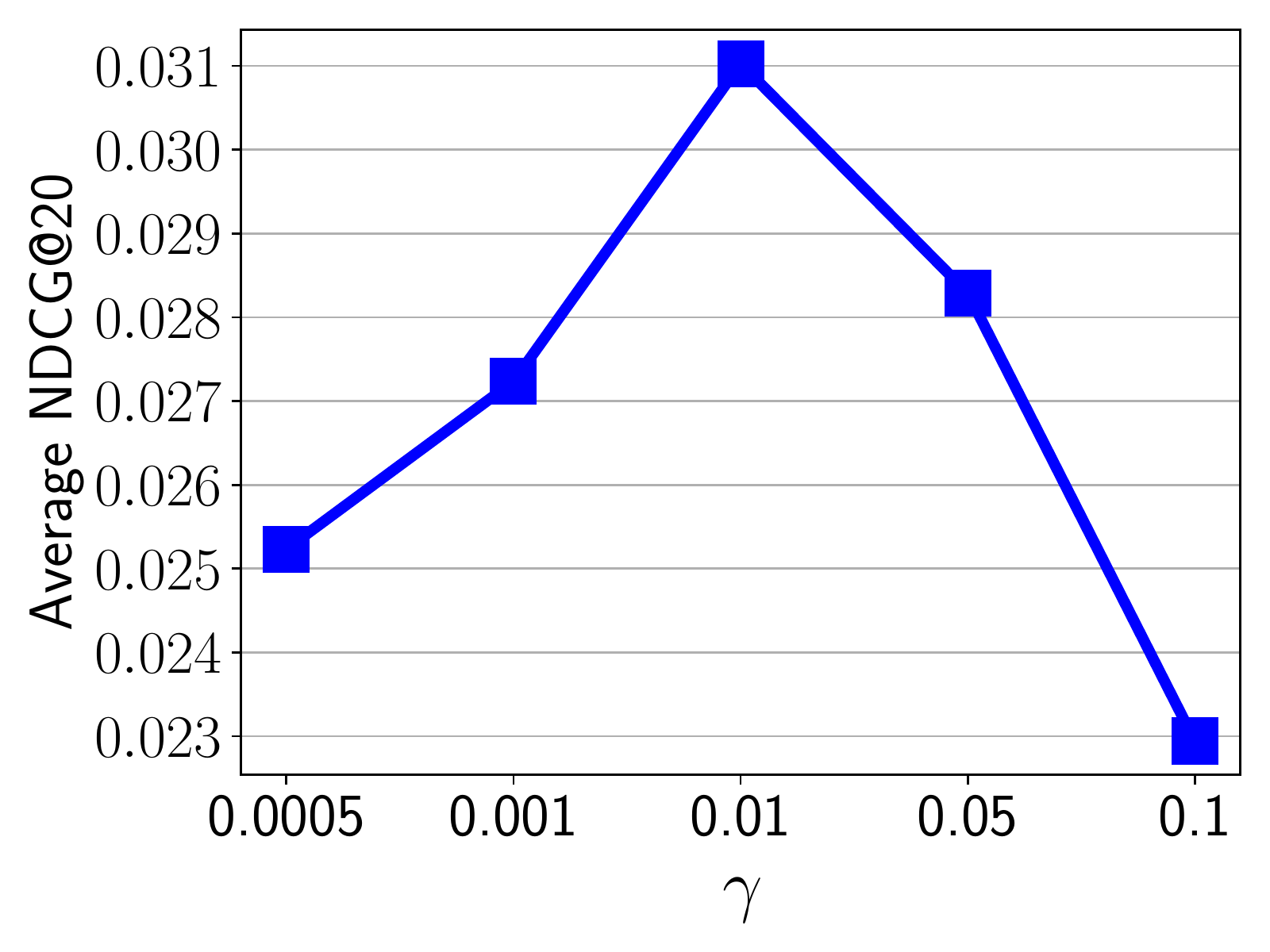}
         \caption{$\gamma$: weight of $\mathcal{L}_{\text{MRA}}$}
         \label{fig:gamma_sens}
     \end{subfigure}
\caption{Item-based recommendation performance sensitivity of each component in Eq.~(\ref{eq:final_loss}).}
\label{fig:sens}
\end{figure}


\begin{table}[]
\centering
\caption{Effects of Pre-training Tasks. }
\label{tab:remove_module}
\begin{tabular}{@{}l|cc|cc@{}}
\toprule
 & \multicolumn{2}{c|}{Knowledge Pred.} & \multicolumn{2}{c}{ZSIR} \\ \hline
Variant & MRR & Recall@20 & MRR & NDCG@20 \\ \hline
\modelname & 0.0142 & 0.0255 & 0.0253 & 0.0310 \\
w/o KR & 0.0041 & 0.0097 & 0.0124 & 0.0152 \\
w/o FR & 0.0122 & 0.0224 & 0.0216 & 0.0242 \\
w/o HNR & 0.0135 & 0.0238 & 0.0234 & 0.0269 \\
w/o MRA & 0.0120 & 0.0217 & 0.0200 & 0.0245 \\ \bottomrule
\end{tabular}%
\end{table}

\subsection{Pre-training Tasks Study~(RQ3)}
\subsubsection{Impacts of Individual Task}
We first show the effectiveness of each pre-training tasks as in Eq.~(\ref{eq:final_loss}) by removing one from the final loss. 
Recall that we have four pre-training tasks, Knowledeg Reconstruction (KR), High-order Neighbor Reconstruction (HNR), Feature Reconstruction (FR) and Meta Relation Adaptation (MRA).
Our ablation study includes the performance on both knowledge prediction and ZSIR tasks after we remove each pre-training tasks, shown in Table~\ref{tab:remove_module}. 
The ZSIR performance is reported as in the average of all markets.
We observe that performance degrades when we remove each task individually.
This demonstrates that all pre-training taskes are necessary for the satisfactory universal item embeddings learning.
Moreover, among all tasks, we observe that performance decreases more significantly when we remove the KR task. 
The reason is that links between items are the key semantics for PKG. Without KR task, the model fails to characterize the intrinsic semantics. 

We visualize the sensitivity of hyper-parameters for each task loss as in the final loss Eq.~(\ref{eq:final_loss}), shown in Figure~\ref{fig:sens}. 
The best weights are not the same for all components because the loss scales are varying for each loss component.
We also observe that the performance drops more significantly for the k-hop neighbors reconstruction loss weight $\theta$.
\subsubsection{Impacts of K in HNR task}
\begin{figure}[]
\centering
\begin{subfigure}[b]{0.23\textwidth}
         \centering
         \includegraphics[width=1\textwidth]{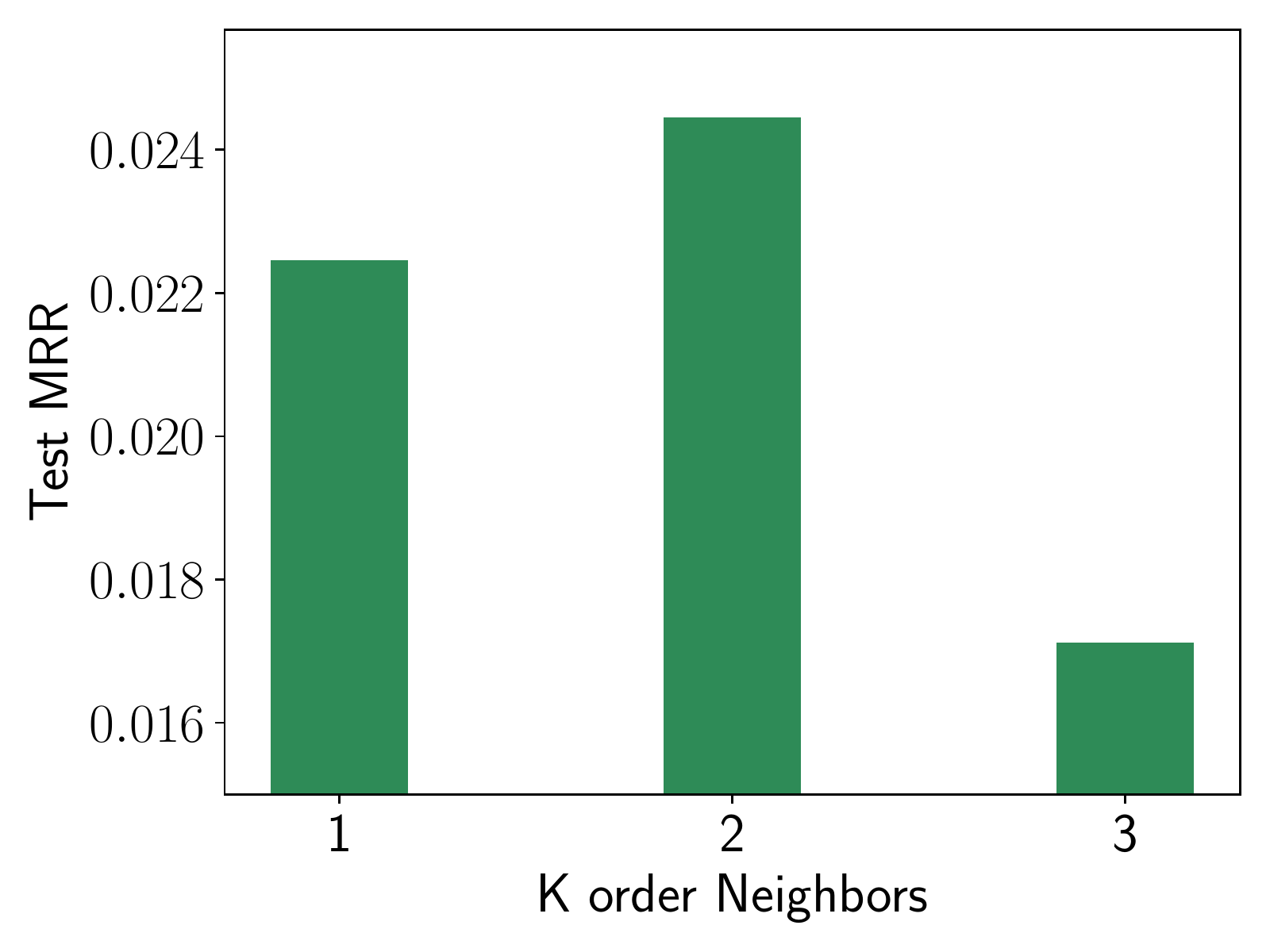}
         \caption{MRR}
         \label{fig:beta_sens}
     \end{subfigure}\hfill
     \begin{subfigure}[b]{0.23\textwidth}
         \centering
         \includegraphics[width=1\textwidth]{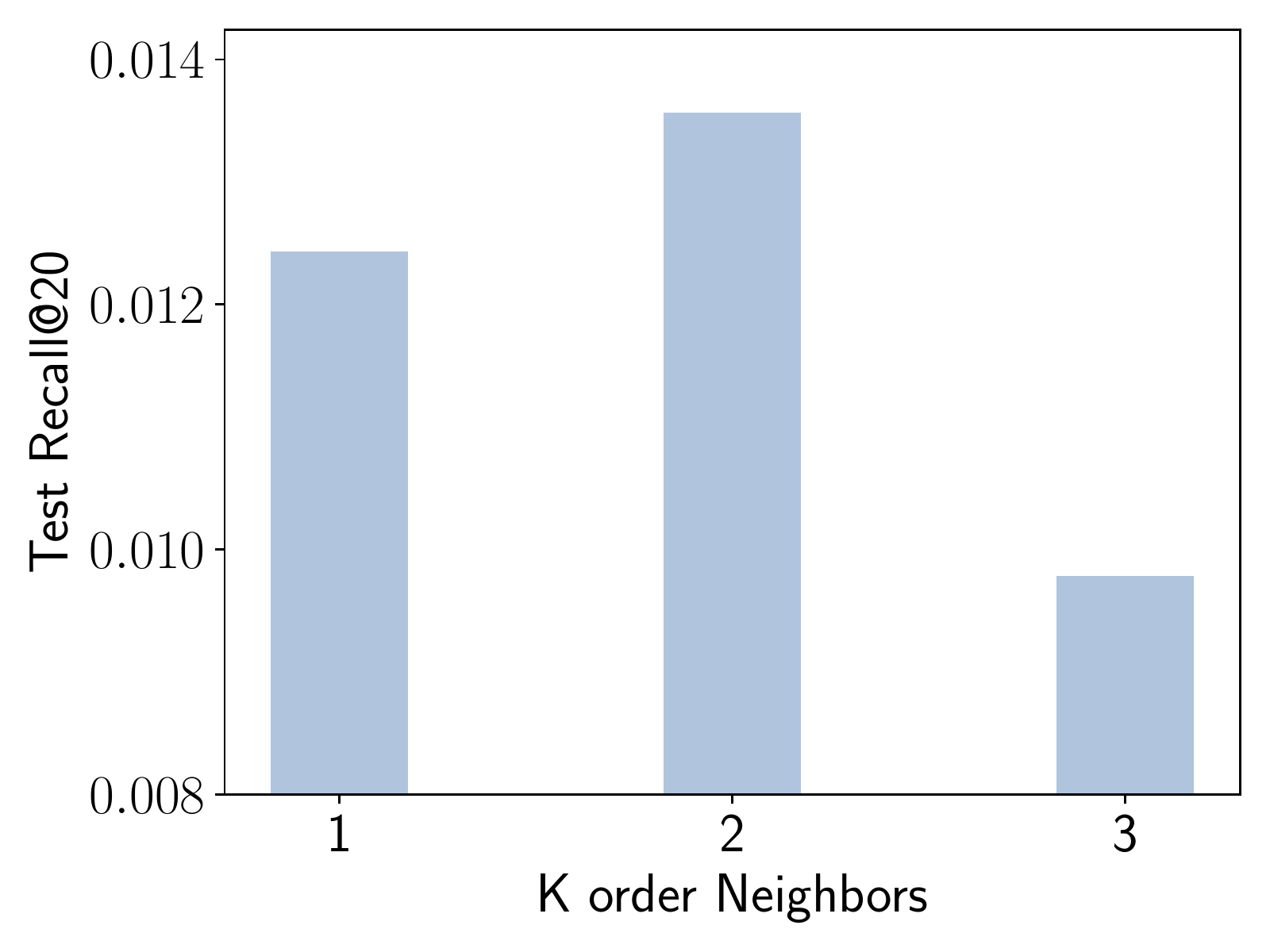}
         \caption{Recall@20}
         \label{fig:khop_sens}
     \end{subfigure}
\caption{Performance Sensitivity of $k$-hop Neighbors on the Knowledge Prediction Task.}
\label{fig:sens_khop_kpred}
\end{figure}

\begin{figure}[]
\centering
\begin{subfigure}[b]{0.23\textwidth}
         \centering
         \includegraphics[width=1\textwidth]{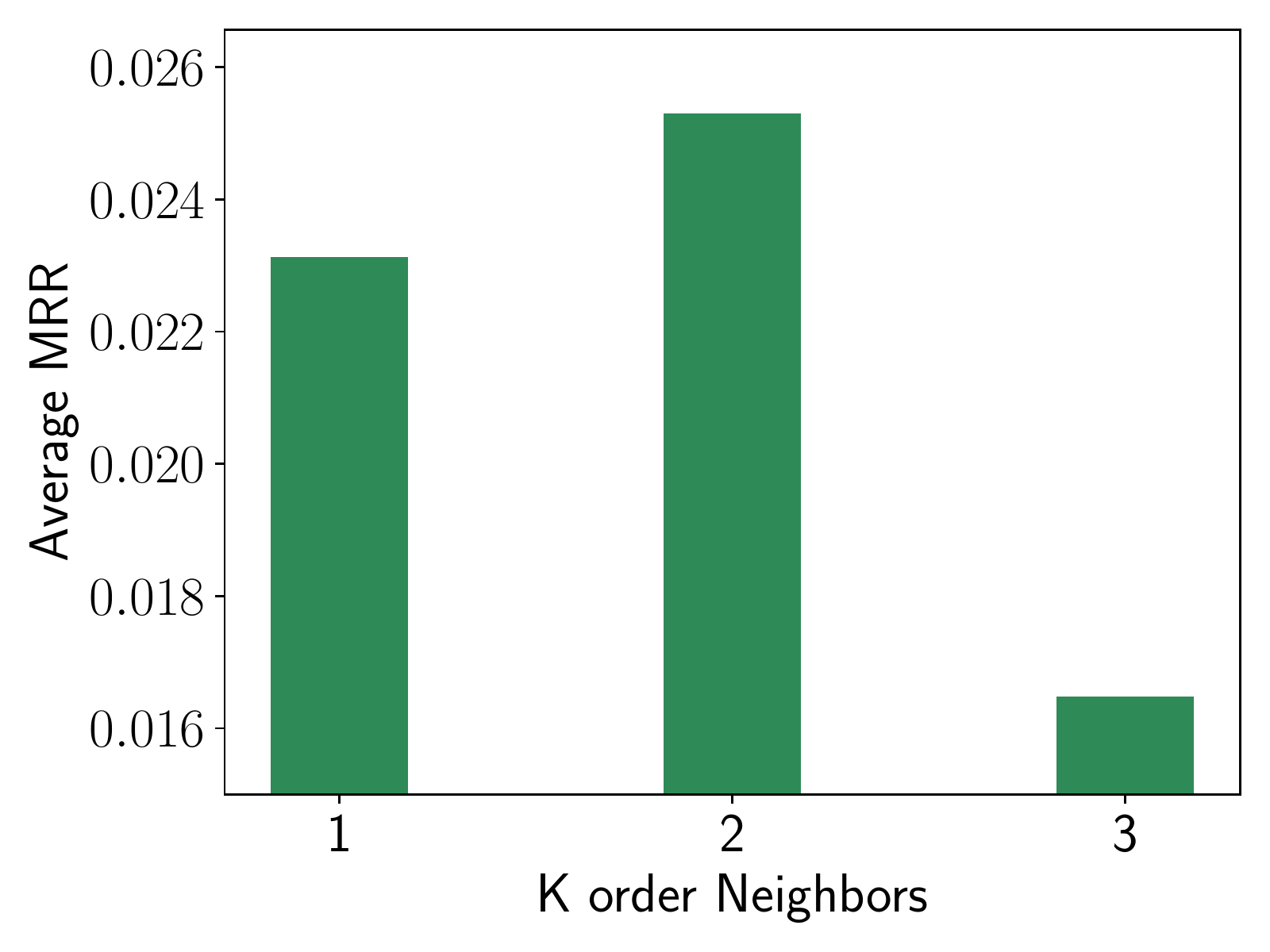}
         \caption{MRR}
         \label{fig:beta_sens}
     \end{subfigure}\hfill
     \begin{subfigure}[b]{0.23\textwidth}
         \centering
         \includegraphics[width=1\textwidth]{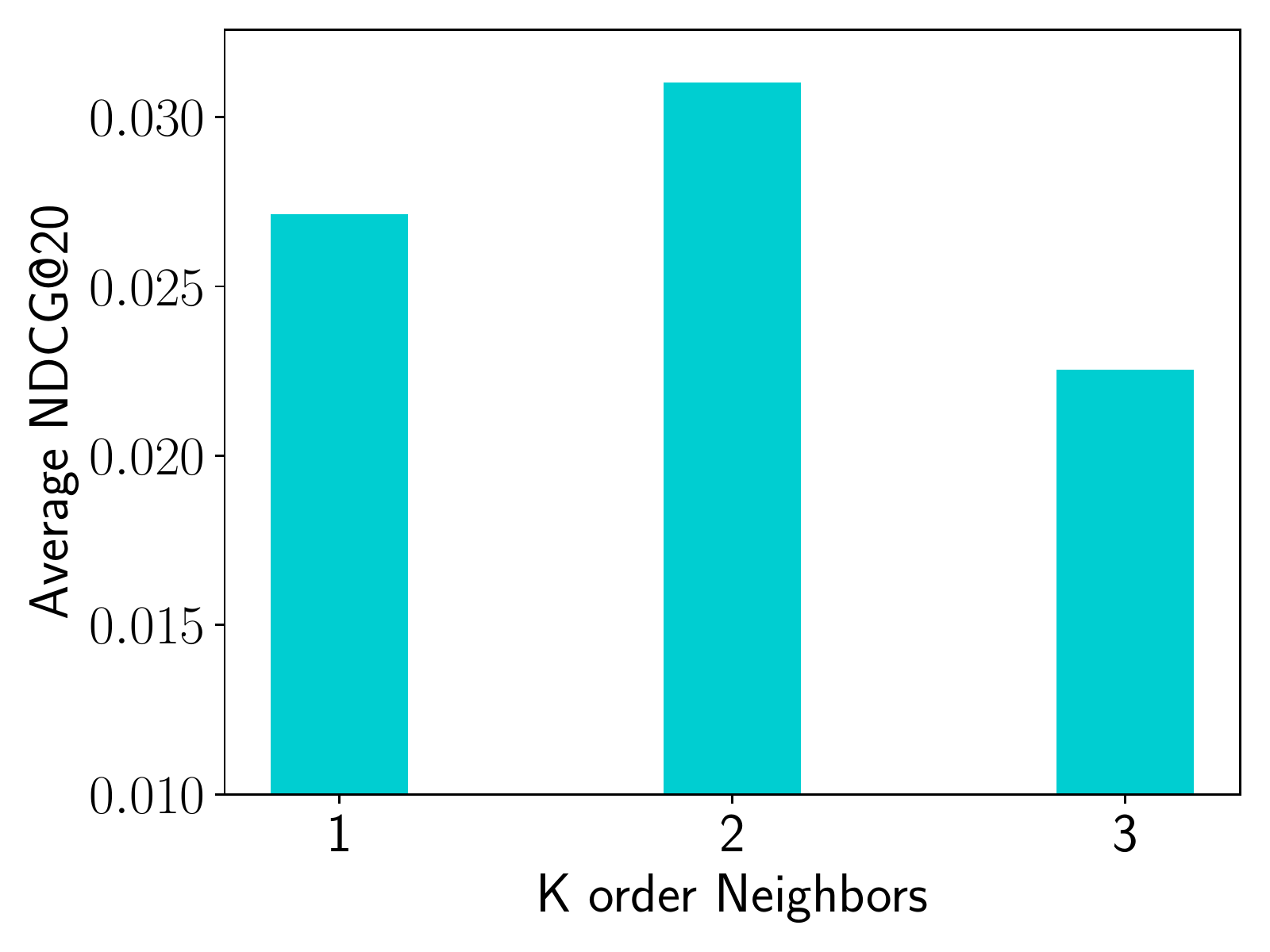}
         \caption{NDCG@20}
         \label{fig:khop_sens}
     \end{subfigure}
\caption{Performance Sensitivity of k-hop Neighbors on the ZSIR task.}
\label{fig:sens_khop_itemrec}
\end{figure}

In this section, we investigate the effect of choosing different $K$s in the k-hop neighbors reconstruction loss $\mathcal{L}_{HNR}$. We choose $K$ from $\{1,2,3\}$ and test it on both tasks. The knowledge prediction task results are shown in Figure~\ref{fig:sens_khop_kpred}, and the item-based recommendation task results are shown in Figure~\ref{fig:sens_khop_itemrec}. From both Figure~\ref{fig:sens_khop_kpred} and Figure~\ref{fig:sens_khop_itemrec}, we can see that when $K=2$, the best performance is achieved. When we include high-order neighbors, such as $K=3$, the performance drops significantly, with the potential reason that high-order neighbors might introduce more irrelevant noises.

\subsection{Effects of Base Model Variants~(RQ4)}
We study the sensitivity of choosing different base models, including the PLM and graph encoder. 
Our proposed framework can adopt arbitrary different GNN encoder and PLM.
The performance sensitivity of different graph encoder variants is shown in Table~\ref{tab:gnn_base}.
For different textual language models that generate universal textual features, we show the results in Table~\ref{tab:llm_variants}. 

Our proposed \modelname adopts the efficient SGC as the base model. 
From Table~\ref{tab:gnn_base}, we observe that the GCN~\cite{welling2016semi} achieves the worst performance. 
The second best GNN encoder is GAT~\cite{veličković2018graph}.
The underlying reason is that SGC is easier to learn and able to generalize to new data.

We also investigated the effects of using different textual language models to generate universal textual features, including Distill-Bert~\cite{sanh2019distilbert}, Bert (adopted in this work)~\cite{devlin2018bert}, and Sentence-Bert~\cite{reimers2019sentence}. In Table~\ref{tab:llm_variants}, we can see that Sentence-Bert achieves marginally better performances than BERT. 
Moreover, Distill-Bert cannot achieve on-par performance, even though its efficiency is significantly better than other two.
\begin{table}[]
\centering
\caption{Effects of Graph Encoder Variants.}
\label{tab:gnn_base}
\begin{tabular}{l|cc|cc}
\toprule
 & \multicolumn{2}{c|}{Knowledge Pred.} & \multicolumn{2}{c}{ZSIR} \\ \hline
GNNs & MRR & Recall@20 & MRR & NDCG@20 \\ \hline
GCN-base & 0.0108 & 0.0231 & 0.0199 & 0.0257 \\
GAT-base & 0.0117 & 0.0235 & 0.0214 & 0.0286 \\
SGC-base & 0.0122 & 0.0241 & 0.0223 & 0.0295 \\ \bottomrule
\end{tabular}%
\end{table}


\begin{table}[]
\centering
\caption{Effects of PLM Variants.}
\label{tab:llm_variants}
\begin{tabular}{l|cc|cc}
\toprule
 & \multicolumn{2}{c|}{Knowledge Pred.} & \multicolumn{2}{c}{ZSIR} \\ \hline
PLMs & MRR & Recall@20 & MRR & NDCG@20 \\ \hline
Distill-Bert & 0.0138 & 0.0238 & 0.0249 & 0.0298 \\
Bert & 0.0142 & 0.0255 & 0.0253 & 0.0310 \\
Sentence-Bert & 0.0145 & 0.0249 & 0.0250 & 0.0304 \\ \bottomrule
\end{tabular}%
\end{table}

\section{Conclusion}
In this work, we investigate the challenging problem of pre-training product knowledge graph to infer universal item representations for zero-shot item-based recommendation. 
We propose four pre-training tasks that comprehensively characterize PKG semantics and improve the adaptation ability of the model to new tasks, including knowledge reconstruction, feature reconstruction, high-order neighbors reconstruction, and the meta relation adaptation tasks. 
We also discuss how to leverage the recommendation task to fine-tune the novel task-oriented adaptation layers such that semantics in PKG can be adapted to new tasks.
Our comprehensive experiments demonstrate the effectiveness of \modelname in capturing existing knowledge for items in PKG via knowledge prediction task and generalization capability in the recommendation task. 
Though in this paper, we only discuss how to fine-tune the model for ZSIR task, our framework is a general pre-training paradiagm for PKG and adaptable to any other new tasks. 



\bibliographystyle{ACM-Reference-Format}
\balance
\bibliography{sample-base}


\begin{thebibliography}{42}


\ifx \showCODEN    \undefined \def \showCODEN     #1{\unskip}     \fi
\ifx \showDOI      \undefined \def \showDOI       #1{#1}\fi
\ifx \showISBNx    \undefined \def \showISBNx     #1{\unskip}     \fi
\ifx \showISBNxiii \undefined \def \showISBNxiii  #1{\unskip}     \fi
\ifx \showISSN     \undefined \def \showISSN      #1{\unskip}     \fi
\ifx \showLCCN     \undefined \def \showLCCN      #1{\unskip}     \fi
\ifx \shownote     \undefined \def \shownote      #1{#1}          \fi
\ifx \showarticletitle \undefined \def \showarticletitle #1{#1}   \fi
\ifx \showURL      \undefined \def \showURL       {\relax}        \fi
\providecommand\bibfield[2]{#2}
\providecommand\bibinfo[2]{#2}
\providecommand\natexlab[1]{#1}
\providecommand\showeprint[2][]{arXiv:#2}

\bibitem[Bonab et~al\mbox{.}(2021)]%
        {bonab2021crossmarket}
\bibfield{author}{\bibinfo{person}{Hamed Bonab}, \bibinfo{person}{Mohammad
  Aliannejadi}, \bibinfo{person}{Ali Vardasbi}, \bibinfo{person}{Evangelos
  Kanoulas}, {and} \bibinfo{person}{James Allan}.}
  \bibinfo{year}{2021}\natexlab{}.
\newblock \showarticletitle{Cross-Market Product Recommendation}. In
  \bibinfo{booktitle}{\emph{Proceedings of the 30th ACM International
  Conference on Information \& Knowledge Management}}.
  \bibinfo{publisher}{ACM}.
\newblock


\bibitem[Bordes et~al\mbox{.}(2013)]%
        {bordes2013translating}
\bibfield{author}{\bibinfo{person}{Antoine Bordes}, \bibinfo{person}{Nicolas
  Usunier}, \bibinfo{person}{Alberto Garcia-Duran}, \bibinfo{person}{Jason
  Weston}, {and} \bibinfo{person}{Oksana Yakhnenko}.}
  \bibinfo{year}{2013}\natexlab{}.
\newblock \showarticletitle{Translating embeddings for modeling
  multi-relational data}.
\newblock \bibinfo{journal}{\emph{Advances in neural information processing
  systems}}  \bibinfo{volume}{26} (\bibinfo{year}{2013}).
\newblock


\bibitem[Chen et~al\mbox{.}(2020)]%
        {chen2020try}
\bibfield{author}{\bibinfo{person}{Tong Chen}, \bibinfo{person}{Hongzhi Yin},
  \bibinfo{person}{Guanhua Ye}, \bibinfo{person}{Zi Huang},
  \bibinfo{person}{Yang Wang}, {and} \bibinfo{person}{Meng Wang}.}
  \bibinfo{year}{2020}\natexlab{}.
\newblock \showarticletitle{Try this instead: Personalized and interpretable
  substitute recommendation}. In \bibinfo{booktitle}{\emph{Proceedings of the
  43rd International ACM SIGIR Conference on Research and Development in
  Information Retrieval}}. \bibinfo{pages}{891--900}.
\newblock


\bibitem[Chen et~al\mbox{.}(2019)]%
        {chen2019personalized}
\bibfield{author}{\bibinfo{person}{Xu Chen}, \bibinfo{person}{Hanxiong Chen},
  \bibinfo{person}{Hongteng Xu}, \bibinfo{person}{Yongfeng Zhang},
  \bibinfo{person}{Yixin Cao}, \bibinfo{person}{Zheng Qin}, {and}
  \bibinfo{person}{Hongyuan Zha}.} \bibinfo{year}{2019}\natexlab{}.
\newblock \showarticletitle{Personalized fashion recommendation with visual
  explanations based on multimodal attention network: Towards visually
  explainable recommendation}. In \bibinfo{booktitle}{\emph{Proceedings of the
  42nd International ACM SIGIR Conference on Research and Development in
  Information Retrieval}}. \bibinfo{pages}{765--774}.
\newblock


\bibitem[Devlin et~al\mbox{.}(2018)]%
        {devlin2018bert}
\bibfield{author}{\bibinfo{person}{Jacob Devlin}, \bibinfo{person}{Ming-Wei
  Chang}, \bibinfo{person}{Kenton Lee}, {and} \bibinfo{person}{Kristina
  Toutanova}.} \bibinfo{year}{2018}\natexlab{}.
\newblock \showarticletitle{Bert: Pre-training of deep bidirectional
  transformers for language understanding}.
\newblock \bibinfo{journal}{\emph{arXiv preprint arXiv:1810.04805}}
  (\bibinfo{year}{2018}).
\newblock


\bibitem[Diao et~al\mbox{.}(2014)]%
        {diao2014jointly}
\bibfield{author}{\bibinfo{person}{Qiming Diao}, \bibinfo{person}{Minghui Qiu},
  \bibinfo{person}{Chao-Yuan Wu}, \bibinfo{person}{Alexander~J Smola},
  \bibinfo{person}{Jing Jiang}, {and} \bibinfo{person}{Chong Wang}.}
  \bibinfo{year}{2014}\natexlab{}.
\newblock \showarticletitle{Jointly modeling aspects, ratings and sentiments
  for movie recommendation (JMARS)}. In \bibinfo{booktitle}{\emph{Proceedings
  of the 20th ACM SIGKDD international conference on Knowledge discovery and
  data mining}}. \bibinfo{pages}{193--202}.
\newblock


\bibitem[Dong et~al\mbox{.}(2020)]%
        {dong2020autoknow}
\bibfield{author}{\bibinfo{person}{Xin~Luna Dong}, \bibinfo{person}{Xiang He},
  \bibinfo{person}{Andrey Kan}, \bibinfo{person}{Xian Li}, \bibinfo{person}{Yan
  Liang}, \bibinfo{person}{Jun Ma}, \bibinfo{person}{Yifan~Ethan Xu},
  \bibinfo{person}{Chenwei Zhang}, \bibinfo{person}{Tong Zhao},
  \bibinfo{person}{Gabriel Blanco~Saldana}, {et~al\mbox{.}}}
  \bibinfo{year}{2020}\natexlab{}.
\newblock \showarticletitle{Autoknow: Self-driving knowledge collection for
  products of thousands of types}. In \bibinfo{booktitle}{\emph{Proceedings of
  the 26th ACM SIGKDD International Conference on Knowledge Discovery \& Data
  Mining}}. \bibinfo{pages}{2724--2734}.
\newblock


\bibitem[Ferrari~Dacrema et~al\mbox{.}(2019)]%
        {ferrari2019we}
\bibfield{author}{\bibinfo{person}{Maurizio Ferrari~Dacrema},
  \bibinfo{person}{Paolo Cremonesi}, {and} \bibinfo{person}{Dietmar Jannach}.}
  \bibinfo{year}{2019}\natexlab{}.
\newblock \showarticletitle{Are we really making much progress? A worrying
  analysis of recent neural recommendation approaches}. In
  \bibinfo{booktitle}{\emph{Proceedings of the 13th ACM conference on
  recommender systems}}. \bibinfo{pages}{101--109}.
\newblock


\bibitem[Geng et~al\mbox{.}(2022)]%
        {geng2022recommendation}
\bibfield{author}{\bibinfo{person}{Shijie Geng}, \bibinfo{person}{Shuchang
  Liu}, \bibinfo{person}{Zuohui Fu}, \bibinfo{person}{Yingqiang Ge}, {and}
  \bibinfo{person}{Yongfeng Zhang}.} \bibinfo{year}{2022}\natexlab{}.
\newblock \showarticletitle{Recommendation as Language Processing (RLP): A
  Unified Pretrain, Personalized Prompt \& Predict Paradigm (P5)}.
\newblock \bibinfo{journal}{\emph{arXiv preprint arXiv:2203.13366}}
  (\bibinfo{year}{2022}).
\newblock


\bibitem[Hamilton et~al\mbox{.}(2017)]%
        {hamilton2017inductive}
\bibfield{author}{\bibinfo{person}{Will Hamilton}, \bibinfo{person}{Zhitao
  Ying}, {and} \bibinfo{person}{Jure Leskovec}.}
  \bibinfo{year}{2017}\natexlab{}.
\newblock \showarticletitle{Inductive representation learning on large graphs}.
\newblock \bibinfo{journal}{\emph{Advances in neural information processing
  systems}}  \bibinfo{volume}{30} (\bibinfo{year}{2017}).
\newblock


\bibitem[He et~al\mbox{.}(2020)]%
        {he2020lightgcn}
\bibfield{author}{\bibinfo{person}{Xiangnan He}, \bibinfo{person}{Kuan Deng},
  \bibinfo{person}{Xiang Wang}, \bibinfo{person}{Yan Li},
  \bibinfo{person}{Yongdong Zhang}, {and} \bibinfo{person}{Meng Wang}.}
  \bibinfo{year}{2020}\natexlab{}.
\newblock \showarticletitle{Lightgcn: Simplifying and powering graph
  convolution network for recommendation}. In
  \bibinfo{booktitle}{\emph{Proceedings of the 43rd International ACM SIGIR
  conference on research and development in Information Retrieval}}.
  \bibinfo{pages}{639--648}.
\newblock


\bibitem[He et~al\mbox{.}(2018)]%
        {he2018nais}
\bibfield{author}{\bibinfo{person}{Xiangnan He}, \bibinfo{person}{Zhankui He},
  \bibinfo{person}{Jingkuan Song}, \bibinfo{person}{Zhenguang Liu},
  \bibinfo{person}{Yu-Gang Jiang}, {and} \bibinfo{person}{Tat-Seng Chua}.}
  \bibinfo{year}{2018}\natexlab{}.
\newblock \showarticletitle{Nais: Neural attentive item similarity model for
  recommendation}.
\newblock \bibinfo{journal}{\emph{IEEE Transactions on Knowledge and Data
  Engineering}} \bibinfo{volume}{30}, \bibinfo{number}{12},
  \bibinfo{pages}{2354--2366}.
\newblock


\bibitem[Hou et~al\mbox{.}(2022)]%
        {hou2022towards}
\bibfield{author}{\bibinfo{person}{Yupeng Hou}, \bibinfo{person}{Shanlei Mu},
  \bibinfo{person}{Wayne~Xin Zhao}, \bibinfo{person}{Yaliang Li},
  \bibinfo{person}{Bolin Ding}, {and} \bibinfo{person}{Ji-Rong Wen}.}
  \bibinfo{year}{2022}\natexlab{}.
\newblock \showarticletitle{Towards Universal Sequence Representation Learning
  for Recommender Systems}. In \bibinfo{booktitle}{\emph{Proceedings of the
  28th ACM SIGKDD Conference on Knowledge Discovery and Data Mining}}.
  \bibinfo{pages}{585--593}.
\newblock


\bibitem[Hu et~al\mbox{.}(2018)]%
        {hu2018squeeze}
\bibfield{author}{\bibinfo{person}{Jie Hu}, \bibinfo{person}{Li Shen}, {and}
  \bibinfo{person}{Gang Sun}.} \bibinfo{year}{2018}\natexlab{}.
\newblock \showarticletitle{Squeeze-and-excitation networks}. In
  \bibinfo{booktitle}{\emph{Proceedings of the IEEE conference on computer
  vision and pattern recognition}}. \bibinfo{pages}{7132--7141}.
\newblock


\bibitem[Hu et~al\mbox{.}(2020)]%
        {hu2020gpt}
\bibfield{author}{\bibinfo{person}{Ziniu Hu}, \bibinfo{person}{Yuxiao Dong},
  \bibinfo{person}{Kuansan Wang}, \bibinfo{person}{Kai-Wei Chang}, {and}
  \bibinfo{person}{Yizhou Sun}.} \bibinfo{year}{2020}\natexlab{}.
\newblock \showarticletitle{Gpt-gnn: Generative pre-training of graph neural
  networks}. In \bibinfo{booktitle}{\emph{Proceedings of the 26th ACM SIGKDD
  International Conference on Knowledge Discovery \& Data Mining}}.
  \bibinfo{pages}{1857--1867}.
\newblock


\bibitem[Huan et~al\mbox{.}(2022)]%
        {huan2022industrial}
\bibfield{author}{\bibinfo{person}{Zhaoxin Huan}, \bibinfo{person}{Gongduo
  Zhang}, \bibinfo{person}{Xiaolu Zhang}, \bibinfo{person}{Jun Zhou},
  \bibinfo{person}{Qintong Wu}, \bibinfo{person}{Lihong Gu},
  \bibinfo{person}{Jinjie Gu}, \bibinfo{person}{Yong He}, \bibinfo{person}{Yue
  Zhu}, {and} \bibinfo{person}{Linjian Mo}.} \bibinfo{year}{2022}\natexlab{}.
\newblock \showarticletitle{An Industrial Framework for Cold-Start
  Recommendation in Zero-Shot Scenarios}. In
  \bibinfo{booktitle}{\emph{Proceedings of the 45th International ACM SIGIR
  Conference on Research and Development in Information Retrieval}}.
  \bibinfo{pages}{3403--3407}.
\newblock


\bibitem[Ji et~al\mbox{.}(2015)]%
        {ji2015knowledge}
\bibfield{author}{\bibinfo{person}{Guoliang Ji}, \bibinfo{person}{Shizhu He},
  \bibinfo{person}{Liheng Xu}, \bibinfo{person}{Kang Liu}, {and}
  \bibinfo{person}{Jun Zhao}.} \bibinfo{year}{2015}\natexlab{}.
\newblock \showarticletitle{Knowledge graph embedding via dynamic mapping
  matrix}. In \bibinfo{booktitle}{\emph{Proceedings of the 53rd annual meeting
  of the association for computational linguistics and the 7th international
  joint conference on natural language processing (volume 1: Long papers)}}.
  \bibinfo{pages}{687--696}.
\newblock


\bibitem[Li et~al\mbox{.}(2019)]%
        {li2019zero}
\bibfield{author}{\bibinfo{person}{Jingjing Li}, \bibinfo{person}{Mengmeng
  Jing}, \bibinfo{person}{Ke Lu}, \bibinfo{person}{Lei Zhu},
  \bibinfo{person}{Yang Yang}, {and} \bibinfo{person}{Zi Huang}.}
  \bibinfo{year}{2019}\natexlab{}.
\newblock \showarticletitle{From zero-shot learning to cold-start
  recommendation}. In \bibinfo{booktitle}{\emph{Proceedings of the AAAI
  conference on artificial intelligence}}, Vol.~\bibinfo{volume}{33}.
  \bibinfo{pages}{4189--4196}.
\newblock


\bibitem[Lu et~al\mbox{.}(2021)]%
        {lu2021learning}
\bibfield{author}{\bibinfo{person}{Yuanfu Lu}, \bibinfo{person}{Xunqiang
  Jiang}, \bibinfo{person}{Yuan Fang}, {and} \bibinfo{person}{Chuan Shi}.}
  \bibinfo{year}{2021}\natexlab{}.
\newblock \showarticletitle{Learning to pre-train graph neural networks}. In
  \bibinfo{booktitle}{\emph{Proceedings of the AAAI Conference on Artificial
  Intelligence}}, Vol.~\bibinfo{volume}{35}. \bibinfo{pages}{4276--4284}.
\newblock


\bibitem[McAuley et~al\mbox{.}(2015)]%
        {mcauley2015inferring}
\bibfield{author}{\bibinfo{person}{Julian McAuley}, \bibinfo{person}{Rahul
  Pandey}, {and} \bibinfo{person}{Jure Leskovec}.}
  \bibinfo{year}{2015}\natexlab{}.
\newblock \showarticletitle{Inferring networks of substitutable and
  complementary products}. In \bibinfo{booktitle}{\emph{Proceedings of the 21th
  ACM SIGKDD international conference on knowledge discovery and data mining}}.
  \bibinfo{pages}{785--794}.
\newblock


\bibitem[Ning and Karypis(2012)]%
        {ning2012sparse}
\bibfield{author}{\bibinfo{person}{Xia Ning} {and} \bibinfo{person}{George
  Karypis}.} \bibinfo{year}{2012}\natexlab{}.
\newblock \showarticletitle{Sparse linear methods with side information for
  top-n recommendations}. In \bibinfo{booktitle}{\emph{Proceedings of the sixth
  ACM conference on Recommender systems}}. \bibinfo{pages}{155--162}.
\newblock


\bibitem[Qiu et~al\mbox{.}(2020)]%
        {qiu2020gcc}
\bibfield{author}{\bibinfo{person}{Jiezhong Qiu}, \bibinfo{person}{Qibin Chen},
  \bibinfo{person}{Yuxiao Dong}, \bibinfo{person}{Jing Zhang},
  \bibinfo{person}{Hongxia Yang}, \bibinfo{person}{Ming Ding},
  \bibinfo{person}{Kuansan Wang}, {and} \bibinfo{person}{Jie Tang}.}
  \bibinfo{year}{2020}\natexlab{}.
\newblock \showarticletitle{Gcc: Graph contrastive coding for graph neural
  network pre-training}. In \bibinfo{booktitle}{\emph{Proceedings of the 26th
  ACM SIGKDD International Conference on Knowledge Discovery \& Data Mining}}.
  \bibinfo{pages}{1150--1160}.
\newblock


\bibitem[Raffel et~al\mbox{.}(2020)]%
        {raffel2020exploring}
\bibfield{author}{\bibinfo{person}{Colin Raffel}, \bibinfo{person}{Noam
  Shazeer}, \bibinfo{person}{Adam Roberts}, \bibinfo{person}{Katherine Lee},
  \bibinfo{person}{Sharan Narang}, \bibinfo{person}{Michael Matena},
  \bibinfo{person}{Yanqi Zhou}, \bibinfo{person}{Wei Li}, {and}
  \bibinfo{person}{Peter~J Liu}.} \bibinfo{year}{2020}\natexlab{}.
\newblock \showarticletitle{Exploring the limits of transfer learning with a
  unified text-to-text transformer}.
\newblock \bibinfo{journal}{\emph{The Journal of Machine Learning Research}}
  \bibinfo{volume}{21}, \bibinfo{number}{1} (\bibinfo{year}{2020}),
  \bibinfo{pages}{5485--5551}.
\newblock


\bibitem[Reimers and Gurevych(2019)]%
        {reimers2019sentence}
\bibfield{author}{\bibinfo{person}{Nils Reimers} {and} \bibinfo{person}{Iryna
  Gurevych}.} \bibinfo{year}{2019}\natexlab{}.
\newblock \showarticletitle{Sentence-bert: Sentence embeddings using siamese
  bert-networks}.
\newblock \bibinfo{journal}{\emph{arXiv preprint arXiv:1908.10084}}
  (\bibinfo{year}{2019}).
\newblock


\bibitem[Rendle et~al\mbox{.}(2012)]%
        {rendle2012bpr}
\bibfield{author}{\bibinfo{person}{Steffen Rendle}, \bibinfo{person}{Christoph
  Freudenthaler}, \bibinfo{person}{Zeno Gantner}, {and} \bibinfo{person}{Lars
  Schmidt-Thieme}.} \bibinfo{year}{2012}\natexlab{}.
\newblock \showarticletitle{BPR: Bayesian personalized ranking from implicit
  feedback}.
\newblock \bibinfo{journal}{\emph{arXiv preprint arXiv:1205.2618}}
  (\bibinfo{year}{2012}).
\newblock


\bibitem[Sanh et~al\mbox{.}(2019)]%
        {sanh2019distilbert}
\bibfield{author}{\bibinfo{person}{Victor Sanh}, \bibinfo{person}{Lysandre
  Debut}, \bibinfo{person}{Julien Chaumond}, {and} \bibinfo{person}{Thomas
  Wolf}.} \bibinfo{year}{2019}\natexlab{}.
\newblock \showarticletitle{DistilBERT, a distilled version of BERT: smaller,
  faster, cheaper and lighter}.
\newblock \bibinfo{journal}{\emph{arXiv preprint arXiv:1910.01108}}
  (\bibinfo{year}{2019}).
\newblock


\bibitem[Veličković et~al\mbox{.}(2018)]%
        {veličković2018graph}
\bibfield{author}{\bibinfo{person}{Petar Veličković},
  \bibinfo{person}{Guillem Cucurull}, \bibinfo{person}{Arantxa Casanova},
  \bibinfo{person}{Adriana Romero}, \bibinfo{person}{Pietro Liò}, {and}
  \bibinfo{person}{Yoshua Bengio}.} \bibinfo{year}{2018}\natexlab{}.
\newblock \showarticletitle{Graph Attention Networks}. In
  \bibinfo{booktitle}{\emph{International Conference on Learning
  Representations}}.
\newblock
\urldef\tempurl%
\url{https://openreview.net/forum?id=rJXMpikCZ}
\showURL{%
\tempurl}


\bibitem[Wang et~al\mbox{.}(2020)]%
        {wang2020make}
\bibfield{author}{\bibinfo{person}{Chenyang Wang}, \bibinfo{person}{Min Zhang},
  \bibinfo{person}{Weizhi Ma}, \bibinfo{person}{Yiqun Liu}, {and}
  \bibinfo{person}{Shaoping Ma}.} \bibinfo{year}{2020}\natexlab{}.
\newblock \showarticletitle{Make it a chorus: knowledge-and time-aware item
  modeling for sequential recommendation}. In
  \bibinfo{booktitle}{\emph{Proceedings of the 43rd International ACM SIGIR
  conference on research and development in Information Retrieval}}.
  \bibinfo{pages}{109--118}.
\newblock


\bibitem[Wang et~al\mbox{.}(2019)]%
        {wang2019neural}
\bibfield{author}{\bibinfo{person}{Xiang Wang}, \bibinfo{person}{Xiangnan He},
  \bibinfo{person}{Meng Wang}, \bibinfo{person}{Fuli Feng}, {and}
  \bibinfo{person}{Tat-Seng Chua}.} \bibinfo{year}{2019}\natexlab{}.
\newblock \showarticletitle{Neural graph collaborative filtering}. In
  \bibinfo{booktitle}{\emph{Proceedings of the 42nd international ACM SIGIR
  conference on Research and development in Information Retrieval}}.
  \bibinfo{pages}{165--174}.
\newblock


\bibitem[Wang et~al\mbox{.}(2021)]%
        {wang2021self}
\bibfield{author}{\bibinfo{person}{Xiao Wang}, \bibinfo{person}{Nian Liu},
  \bibinfo{person}{Hui Han}, {and} \bibinfo{person}{Chuan Shi}.}
  \bibinfo{year}{2021}\natexlab{}.
\newblock \showarticletitle{Self-supervised heterogeneous graph neural network
  with co-contrastive learning}. In \bibinfo{booktitle}{\emph{Proceedings of
  the 27th ACM SIGKDD Conference on Knowledge Discovery \& Data Mining}}.
  \bibinfo{pages}{1726--1736}.
\newblock


\bibitem[Wang et~al\mbox{.}(2014)]%
        {wang2014knowledge}
\bibfield{author}{\bibinfo{person}{Zhen Wang}, \bibinfo{person}{Jianwen Zhang},
  \bibinfo{person}{Jianlin Feng}, {and} \bibinfo{person}{Zheng Chen}.}
  \bibinfo{year}{2014}\natexlab{}.
\newblock \showarticletitle{Knowledge graph embedding by translating on
  hyperplanes}. In \bibinfo{booktitle}{\emph{Proceedings of the AAAI conference
  on artificial intelligence}}, Vol.~\bibinfo{volume}{28}.
\newblock


\bibitem[Wei et~al\mbox{.}(2022)]%
        {wei2022contrastive}
\bibfield{author}{\bibinfo{person}{Wei Wei}, \bibinfo{person}{Chao Huang},
  \bibinfo{person}{Lianghao Xia}, \bibinfo{person}{Yong Xu},
  \bibinfo{person}{Jiashu Zhao}, {and} \bibinfo{person}{Dawei Yin}.}
  \bibinfo{year}{2022}\natexlab{}.
\newblock \showarticletitle{Contrastive meta learning with behavior
  multiplicity for recommendation}. In \bibinfo{booktitle}{\emph{Proceedings of
  the Fifteenth ACM International Conference on Web Search and Data Mining}}.
  \bibinfo{pages}{1120--1128}.
\newblock


\bibitem[Welling and Kipf(2016)]%
        {welling2016semi}
\bibfield{author}{\bibinfo{person}{Max Welling} {and} \bibinfo{person}{Thomas~N
  Kipf}.} \bibinfo{year}{2016}\natexlab{}.
\newblock \showarticletitle{Semi-supervised classification with graph
  convolutional networks}. In \bibinfo{booktitle}{\emph{J. International
  Conference on Learning Representations (ICLR 2017)}}.
\newblock


\bibitem[Wu et~al\mbox{.}(2019)]%
        {wu2019simplifying}
\bibfield{author}{\bibinfo{person}{Felix Wu}, \bibinfo{person}{Amauri Souza},
  \bibinfo{person}{Tianyi Zhang}, \bibinfo{person}{Christopher Fifty},
  \bibinfo{person}{Tao Yu}, {and} \bibinfo{person}{Kilian Weinberger}.}
  \bibinfo{year}{2019}\natexlab{}.
\newblock \showarticletitle{Simplifying graph convolutional networks}. In
  \bibinfo{booktitle}{\emph{International conference on machine learning}}.
  PMLR, \bibinfo{pages}{6861--6871}.
\newblock


\bibitem[Xu et~al\mbox{.}(2020a)]%
        {xu2020knowledge}
\bibfield{author}{\bibinfo{person}{Da Xu}, \bibinfo{person}{Chuanwei Ruan},
  \bibinfo{person}{Jason Cho}, \bibinfo{person}{Evren Korpeoglu},
  \bibinfo{person}{Sushant Kumar}, {and} \bibinfo{person}{Kannan Achan}.}
  \bibinfo{year}{2020}\natexlab{a}.
\newblock \showarticletitle{Knowledge-aware complementary product
  representation learning}. In \bibinfo{booktitle}{\emph{Proceedings of the
  13th International Conference on Web Search and Data Mining}}.
  \bibinfo{pages}{681--689}.
\newblock


\bibitem[Xu et~al\mbox{.}(2020b)]%
        {xu2020inductive}
\bibfield{author}{\bibinfo{person}{Da Xu}, \bibinfo{person}{Chuanwei Ruan},
  \bibinfo{person}{Evren Korpeoglu}, \bibinfo{person}{Sushant Kumar}, {and}
  \bibinfo{person}{Kannan Achan}.} \bibinfo{year}{2020}\natexlab{b}.
\newblock \showarticletitle{Inductive representation learning on temporal
  graphs}.
\newblock \bibinfo{journal}{\emph{arXiv preprint arXiv:2002.07962}}
  (\bibinfo{year}{2020}).
\newblock


\bibitem[Xu et~al\mbox{.}(2020c)]%
        {xu2020product}
\bibfield{author}{\bibinfo{person}{Da Xu}, \bibinfo{person}{Chuanwei Ruan},
  \bibinfo{person}{Evren Korpeoglu}, \bibinfo{person}{Sushant Kumar}, {and}
  \bibinfo{person}{Kannan Achan}.} \bibinfo{year}{2020}\natexlab{c}.
\newblock \showarticletitle{Product knowledge graph embedding for e-commerce}.
  In \bibinfo{booktitle}{\emph{Proceedings of the 13th international conference
  on web search and data mining}}. \bibinfo{pages}{672--680}.
\newblock


\bibitem[Xue et~al\mbox{.}(2019)]%
        {xue2019deep}
\bibfield{author}{\bibinfo{person}{Feng Xue}, \bibinfo{person}{Xiangnan He},
  \bibinfo{person}{Xiang Wang}, \bibinfo{person}{Jiandong Xu},
  \bibinfo{person}{Kai Liu}, {and} \bibinfo{person}{Richang Hong}.}
  \bibinfo{year}{2019}\natexlab{}.
\newblock \showarticletitle{Deep item-based collaborative filtering for top-n
  recommendation}.
\newblock \bibinfo{journal}{\emph{ACM Transactions on Information Systems
  (TOIS)}} \bibinfo{volume}{37}, \bibinfo{number}{3}, \bibinfo{pages}{1--25}.
\newblock


\bibitem[Yan et~al\mbox{.}(2022)]%
        {yan2022personalized}
\bibfield{author}{\bibinfo{person}{An Yan}, \bibinfo{person}{Chaosheng Dong},
  \bibinfo{person}{Yan Gao}, \bibinfo{person}{Jinmiao Fu},
  \bibinfo{person}{Tong Zhao}, \bibinfo{person}{Yi Sun}, {and}
  \bibinfo{person}{Julian McAuley}.} \bibinfo{year}{2022}\natexlab{}.
\newblock \showarticletitle{Personalized complementary product recommendation}.
  In \bibinfo{booktitle}{\emph{Companion Proceedings of the Web Conference
  2022}}. \bibinfo{pages}{146--151}.
\newblock


\bibitem[Zhang et~al\mbox{.}(2018)]%
        {zhang2018knowledge}
\bibfield{author}{\bibinfo{person}{Zhao Zhang}, \bibinfo{person}{Fuzhen
  Zhuang}, \bibinfo{person}{Meng Qu}, \bibinfo{person}{Fen Lin}, {and}
  \bibinfo{person}{Qing He}.} \bibinfo{year}{2018}\natexlab{}.
\newblock \showarticletitle{Knowledge graph embedding with hierarchical
  relation structure}. In \bibinfo{booktitle}{\emph{Proceedings of the 2018
  Conference on Empirical Methods in Natural Language Processing}}.
  \bibinfo{pages}{3198--3207}.
\newblock


\bibitem[Zhao et~al\mbox{.}(2017)]%
        {zhao2017improving}
\bibfield{author}{\bibinfo{person}{Tong Zhao}, \bibinfo{person}{Julian
  McAuley}, \bibinfo{person}{Mengya Li}, {and} \bibinfo{person}{Irwin King}.}
  \bibinfo{year}{2017}\natexlab{}.
\newblock \showarticletitle{Improving recommendation accuracy using networks of
  substitutable and complementary products}. In \bibinfo{booktitle}{\emph{2017
  International Joint Conference on Neural Networks (IJCNN)}}. IEEE,
  \bibinfo{pages}{3649--3655}.
\newblock


\bibitem[Zhou et~al\mbox{.}(2022)]%
        {zhou2022decoupled}
\bibfield{author}{\bibinfo{person}{Zhiheng Zhou}, \bibinfo{person}{Tao Wang},
  \bibinfo{person}{Linfang Hou}, \bibinfo{person}{Xinyuan Zhou},
  \bibinfo{person}{Mian Ma}, {and} \bibinfo{person}{Zhuoye Ding}.}
  \bibinfo{year}{2022}\natexlab{}.
\newblock \showarticletitle{Decoupled Hyperbolic Graph Attention Network for
  Modeling Substitutable and Complementary Item Relationships}. In
  \bibinfo{booktitle}{\emph{Proceedings of the 31st ACM International
  Conference on Information \& Knowledge Management}}.
  \bibinfo{pages}{2763--2772}.
\newblock


\end{thebibliography}

\end{document}